\documentclass[10pt]{article}
\usepackage{subfig}
\usepackage{xcolor}
\usepackage{graphicx}
\usepackage{url}
\usepackage{amsmath,amsfonts,amssymb}
 
\newcommand{\new}[1]{\textcolor{black}{#1}} 
\newcommand{\HMM}{HMs}

\title{\textbf{A Comparative Study of Human Motion Models in Reinforcement Learning Algorithms for Social Robot Navigation}}

\author{Tommaso Van Der Meer, Andrea Garulli, Antonio Giannitrapani, Renato Quartullo}
\date{Dipartimento di Ingegneria dell'Informazione e Scienze Matematiche,\\Università di Siena, Italy}

\begin{document}

\maketitle

\begin{abstract}
Social robot navigation is an evolving research field that aims to find efficient strategies to safely navigate dynamic environments populated by humans. A critical challenge in this domain is the accurate modeling of human motion, which directly impacts the design and evaluation of navigation algorithms. This paper presents a comparative study of two popular categories of human motion models used in social robot navigation, namely velocity-based models and force-based models. A system-theoretic representation of both model types is presented, which highlights their common feedback structure, although with different state variables. Several navigation policies based on reinforcement learning are trained and tested in various simulated environments involving pedestrian crowds modeled with these approaches. A comparative study is conducted to assess performance across multiple factors, including human motion model, navigation policy, scenario complexity and crowd density. The results highlight advantages and challenges of different approaches to modeling human behavior, as well as their role during training and testing of learning-based navigation policies. The findings offer valuable insights and guidelines for selecting appropriate human motion models when designing socially-aware robot navigation systems. 
\end{abstract}

\section{Introduction}
Social robot navigation is a research area that is attracting an increasing attention from researchers of different fields. This is testified by the steadily growing number of publications on the subject, including several recent surveys, see, e.g.,  \cite{charalampous2017recent,cheng2018autonomous,moller2021survey,medina2022perception,mavrogiannis2023core,singamaneni2024survey,karwowski2024bridging}.
Among the many challenges that arise in the design process of a socially-aware navigation pipeline, a key role is played by modeling human motion. 
A mobile robot navigating a crowded environment must typically pursue two requirements, which may be conflicting: reach the assigned goal in an efficient and effective way; choose trajectories that are safe and comfortable for the surrounding humans.  In order to design navigation algorithms which are able to achieve these objectives, reliable models describing human motion are used to train and test navigation policies in different scenarios. As a consequence, both the navigation algorithms and the evaluation of their performance will depend heavily on the chosen human motion model.

In \cite{mavrogiannis2023core}, a detailed taxonomy of crowd dynamic models is presented, based on the level of detail that the model intend to achieve. While macroscopic and mesoscopic models are mainly used to simulate high-level behaviors, such as crowd flow or group formation, microscopic models aim at representing the motion of every individual in the crowd. Therefore, they are more suitable to address key issues in social navigation, such as reciprocal avoidance between robots and humans or prediction of trajectory occlusions due to potential path crossings. In turn, microscopic models can be coarsely divided in velocity-based and force-based models.
The former usually design the agent trajectories on the basis of the observed velocities of neighboring agents. Popular versions are those based on the concept of reciprocal velocity obstacle \cite{van2008reciprocal}, and extended versions employing the more general acceleration-velocity obstacle, like the widely used Optimal Reciprocal Collision Avoidance (ORCA) model \cite{van2011nbody,van2011reciprocal}. A recently proposed velocity-based model is the Social Momentum \cite{mavrogiannis2022social}, in which the robot action is selected so as to trade-off progress of the agent to the goal and avoidance of humans. Conversely, force-based models represent pedestrians as mass particles subject to attractive and repulsive forces, modeling respectively the desire to proceed towards a target location and the tendency to avoid hitting other humans or obstacles. The original Social Force Model (SFM) \cite{helbing2000simulating} and its subsequent variants (see, e.g., \cite{moussaid2009experimental,guo2014simulation}) feature agents that can move equivalently in every direction. The Headed Social Force Model (HSFM) \cite{HSFM} takes into account also the heading of humans, thus generating more realistic behaviors (for example, limiting sideways or backwards movements). 
Intuitively, velocity-based models are explicitly designed to favor trajectory efficiency, while force-based models tend to enforce collision avoidance, possibly at the expense of path optimality.
\new{Many alternative approaches have been proposed in the literature to describe crowd behaviors. For instance, a rich research line concerns the use of data-driven architectures for predicting trajectories of humans in crowds (see, e.g., \cite{alahi2016social,vemula2018social,salzmann2020trajectron++}). The interested reader is referred to \cite{rudenko2020human} for a comprehensive survey of prediction techniques for human motion. On the other hand, it is also worth mentioning recent approaches that address social navigation without requiring an explicit human motion model \cite{cathcart2023proactive}. In this work, we restrict our attention to velocity-based and force-based motion models because, despite they are simple to implement, they are able to express crowd behaviors that are rich enough for designing effective robot navigation techniques.}

The aforementioned models are widely employed in the literature, in both the training and the testing phase of the navigation design process. In this work, we focus on reinforcement learning methods due to their ability to operate in dynamic environments and their adaptability across various scenarios, without the need for manually adjusting parameters as required by model-based approaches, like, e.g., \cite{fox1997dynamic,fiorini1998motion}. Some examples are briefly recalled next. A navigation scheme based on reinforcement learning (referred to as CADRL) is proposed in \cite{chen2017decentralized}, in which human trajectories generated by ORCA are used for the initial training of the robot policy via imitation learning. The approach is then extended in \cite{everett2018motion} to cope with multiple heterogeneous agents, by using long short-term memory networks (LSTM-RL). In \cite{SARL}, a social attentive navigation policy (SARL) is introduced and compared to other navigation techniques based on reinforcement learning, within a simulation environment in which humans move according to ORCA.
The SFM is used in \cite{ferrer2013robot} to design a human-robot interaction scheme for a robot companion. In \cite{tai2018socially}, a SFM-based simulator is employed for training and testing socially-compliant navigation policies based on adversarial imitation learning from raw depth images. 
\new{Socially-aware navigation patterns based on potential field methods and limit cycles are proposed in \cite{boldrer2020socially}.}
A robot planning approach based on model predictive control is proposed in \cite{brito2019model} and tested in an environment with humans moving according to the SFM. In \cite{antonucci2021efficient}, a learning-based approach is presented which embeds the SFM directly into a neural network in order to predict human trajectories. A human motion model combining velocity obstacles and the HSFM is adopted in \cite{fuad2021modified}. Both velocity-based and force-based human motion models are also widely employed within simulation tools and benchmarks for social robot navigation. In particular, ORCA, SFM and their variants are used in popular simulation platforms such as MengeROS \cite{aroor2017mengeros}, CrowdBot \cite{grzeskowiak2021crowd}, HuNavSim \cite{perez2023hunavsim}, SocialGym \cite{chandra2024socialgym}, and many others (see \cite{kaur2022simulators} for a survey on simulators for social robot navigation).

\new{Human motion models are particularly important for learning-based navigation techniques, as these require a large training set that must be generated in simulation before deploying the policy on a real robot. However, the role and influence of human motion models are not yet fully understood.} In  \cite{mavrogiannis2023core} some tests are presented, involving several learning-based navigation policies in environments with humans moving according to either the ORCA or the SFM model. As a major outcome of the study, it is argued that the performance of navigation algorithms strongly depends on the models that are employed to simulate crowd dynamics in both the training and the testing phase.
In particular, the results can change dramatically when such phases are carried out in environments featuring different human motion models.
Another key issue concerns how the navigation performance is assessed. The recent work \cite{francis2023principles} surveys a wide variety of metrics for the evaluation of social robot navigation algorithms. A major classification is the one distinguishing between success metrics, which are more related to the robot navigation task, and social metrics, taking into account human-oriented requirements such as safety and comfort. Clearly, also the computation of such metrics depends on the human motion model adopted in the simulation environment.
A further crucial aspect is the choice of the training and testing scenarios. Usually, social navigation scenarios must be simple enough to limit the complexity of the training phase, but at the same time they must include some of the most common behaviors occurring in the real world, like crossings, parallel traffic, overtaking, etc. (see \cite{francis2023principles} for a detailed classification). It is apparent that the challenges posed by the navigation scenarios depend on the motion models used to simulate human behaviors.

The aim of this paper is to present a comparative study of velocity-based and force-based human motion models for social robot navigation. Specifically, ORCA, SFM and HSFM are considered. \new{The first contribution is a system-theoretic representation of both velocity-based and force-based models. It is shown that such models share a common feedback structure, differing essentially in the state variables chosen to describe the behavior of the human agent.} As a side contribution, an enhanced version of the HSFM able to further promote collision avoidance is presented. Then, three learning-based navigation policies (CADRL, LSTM-RL and SARL) are trained and tested in different scenarios, involving a crowd of pedestrians moving according to the considered motion models. \new{The main contribution is a thorough comparative study that addresses a number of factors, including: the adopted human motion model; the chosen navigation policy; the scenarios in which the policies are trained and tested; the crowd density in the scenario; the effect of noisy measurements on the navigation performance.} Different navigation and social metrics are used to evaluate the performance.
The outcome of the presented comparisons provide useful guidelines for the choice of a suitable human motion model in the context of socially-aware robot navigation.

The rest of the paper is structured as follows.
In Section \ref{human_motion_models}, the human motion models employed in the simulations are presented.
Section \ref{nav_alg} describes the different navigation policies used to guide the robot.
In Section \ref{scenarios}, the scenarios employed during the testing and training phases of this study are presented.
Section \ref{metrics} introduces the metrics used for the evaluations of all trained policies.
In Section \ref{num_results}, the results of an extended simulation campaign are reported and discussed.
In Section \ref{conclusions} some conclusions are drawn and possible future research directions are outlined. The software developed to perform all the simulations is available at: \url{https://github.com/TommasoVandermeer/Social-Navigation-PyEnvs}.


\section{Human motion models}\label{human_motion_models}

One of the main aspects that characterize social navigation is modeling how humans move within the operating environment. In order to design robot navigation policies capable of producing socially acceptable trajectories in the real world, it is crucial to choose appropriate models to reproduce the behavior of a human crowd. 
\new{In this section, a feedback representation of  mathematical models describing human motion is introduced and then specialized to three model classes widely employed in the literature.}

\begin{figure}[t]
  \centering
  \includegraphics[width=0.7\columnwidth]{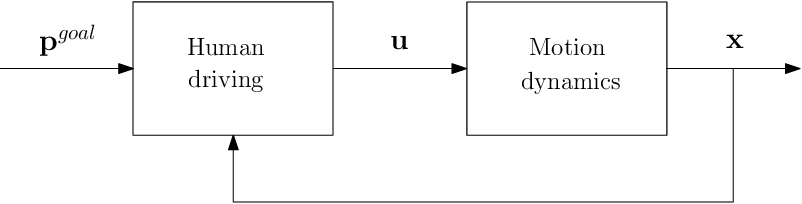}
  \caption{A general scheme for human motion models.}
  \label{fig:hmm}
\end{figure}

A general scheme for human motion models is depicted in Figure~\ref{fig:hmm}. In this work, humans are modeled as agents moving in a two-dimensional environment. Vectors
$\mathbf{p}_{i} = [p_{x},p_{y}]^{T}$ and  $\mathbf{v}_{i} = [v_{x},v_{y}]^{T}$ represent the position and velocity of the $i$-th pedestrian, respectively. The orientation of the agent with respect to the $x$-axis is denoted by $\theta_i$ and their angular velocity by $\omega_i$. The space occupied by the $i$-th pedestrian is a disc of radius $r_i$ centered in $\mathbf{p}_{i}$ and their mass is denoted by $m_i$. Given a group of $n$ humans, the \emph{motion dynamics} block in Figure~\ref{fig:hmm} is described by a set of differential equations 
\begin{equation}\label{eq:motdyn}
\dot{\mathbf{x}} = F( \mathbf{x}, \mathbf{u})
\end{equation}
in which $\mathbf{x}=[\mathbf{x}_1^T,\mathbf{x}_2^T,\dots,\mathbf{x}_n^T]^T$ and $\mathbf{u}=[\mathbf{u}_1^T,\mathbf{u}_2^T,\dots,\mathbf{u}_n^T]^T$ are the state and input  vector of the system, respectively. Vectors $\mathbf{x}_i$ and $\mathbf{u}_i$ denote the state and input of the $i$-th pedestrian, whose definition depends on the specific human motion model class.
The \emph{human driving} block in Figure~\ref{fig:hmm} encodes the way the input of the motion dynamics is selected. It is usually a nonlinear function of the system state and of a reference vector $\mathbf{p}^{goal}$, containing the positions that the pedestrians aim to reach. This feedback term is a specific feature of each human motion model.

In order to simulate the motion of a human crowd, it is sufficient to generate a sequence of waypoints for each pedestrian and feed such sequence to the system as the reference $\mathbf{p}^{goal}$. The human driving term takes care of tracking the reference waypoints, while preventing humans from bumping into each other. In the following, the three different classes of human motion models employed in this work are recalled.

\subsection{Optimal Reciprocal Collision Avoidance}

Optimal Reciprocal Collision Avoidance (ORCA) is a model specifically designed for collision avoidance, which relies on the concept of acceleration-velocity obstacles (AVOs) \cite{van2011reciprocal}, an extension of the classical velocity obstacles (VOs) \cite{fiorini1998motion}.
Given a pair of agents $i$ and $j$, the acceleration-velocity obstacle $AVO_{ij}$ is the set of new relative velocities of agent $i$ with respect to $j$ that will cause a collision between the agents within a prescribed time, if both agents use proportional control of the acceleration to achieve the aforementioned relative velocity. The set $AVO_{ij}$ is a function of the current positions and velocities of agents $i$ and $j$ and closed form expressions for its boundaries are available. 

In ORCA, the agent state is defined by their position and velocity, namely $\mathbf{x}_i=[\mathbf{p}_i^T, \mathbf{v}_i^T]^T$. The input $\mathbf{u}_i$ is the agent acceleration, thus giving the classical double-integrator motion dynamics
\new{
\begin{align*}
\dot{\mathbf{p}}_{i} &= \mathbf{v}_{i}  \\
\dot{\mathbf{v}}_{i}  &= \mathbf{u}_{i}.
\end{align*}
}
The human driving term is designed in such a way to guarantee reciprocal collision avoidance between the agents. This is done by defining the set of feasible velocities $V_i$ for each agent $i$, i.e., the set of all velocities that pedestrian $i$ can choose without colliding with the other humans. Exploiting the notion of AVO, this is given by
$$
V_i= D_i \setminus \bigcup_{j \neq i} \left( AVO_{ij} \oplus \{ \mathbf{v}_j \} \right)
$$
where $D_i$ represents the set of velocities of agent $i$ compatible with physical constraints associated to pedestrian motions (e.g., an upper bound on the acceleration).
Then, the new velocity $\mathbf{v}'_i$ of agent $i$ is selected as the one that is closer to their preferred velocity $\mathbf{v}_i^{d}$, among all the feasible ones, i.e.
\begin{equation}
    \mathbf{v}'_i = \arg \min_{\mathbf{v} \in V_i} ||\mathbf{v} - \mathbf{v}_i^{d}||
\end{equation}
Finally, the acceleration input is chosen as $\mathbf{u}_i=(\mathbf{v}'_{i}-\mathbf{v}_{i})/\delta$, where $\delta$ is a parameter chosen for proportional control of the acceleration.
In this work, the preferred velocity vector $\mathbf{v}_i^{d}$ of agent $i$ is defined as their maximum velocity, denoted by $v_i^{max}$, towards the agent goal, i.e. 
\begin{equation}
    \mathbf{v}_{i}^{d} = v_i^{max} \frac{\mathbf{p}_i^{goal} - \mathbf{p}_{i}}{||\mathbf{p}_i^{goal} - \mathbf{p}_{i}||}.
    \label{eq:desvel}
\end{equation}

\subsection{Social Force Model}

The Social Force Model (SFM) is a human motion model in which each pedestrian is described as a mass particle driven by attractive and repulsive forces  \cite{helbing2000simulating}.
In the SFM setting, the state of agent $i$ contains their position and velocity,  $\mathbf{x}_i=[\mathbf{p}_i^T, \mathbf{v}_i^T]^T$, while the input is the total force acting on the $i$-th mass particle. This results in the motion dynamics equations
\begin{align*}
\dot{\mathbf{p}}_{i} &= \mathbf{v}_{i} \\
\dot{\mathbf{v}}_{i}  &= \frac{1}{m_{i}}\mathbf{u}_{i},
\end{align*}
where $m_i$ is the mass of the $i$-th pedestrian.
The human driving law is defined as
\begin{equation}
    \mathbf{u}_{i} = \mathbf{f}_{i}^{d} + \sum_{j \neq i} \mathbf{f}_{ij}^{p},
\end{equation}
where $\mathbf{f}_{i}^{d}$ represents the so-called desired force, attracting the $i$-th pedestrian towards their goal, and  $\mathbf{f}_{ij}^{p}$ is the repulsive force exerted from the $j$-th to the $i$-th pedestrian, modeling the tendency of humans to stay away from each other. The desired force $\mathbf{f}_{i}^{d}$ is given by 
\begin{equation}\label{eq:fi0}
    \mathbf{f}_{i}^{d} = m_{i} \frac{\mathbf{v}_{i}^{d} - \mathbf{v}_{i}}{\tau_{i}},
\end{equation}
where $\mathbf{v}_{i}^{d}$ is the the preferred velocity of the pedestrian towards the goal, computed as in \eqref{eq:desvel}, and $\tau_{i} > 0$ is the relaxation time which regulates the rate of change of the velocity.
\new{The smaller is $\tau_{i}$, the faster the $i$-th agent will adapt its current velocity $\mathbf{v}_{i}$ to the desired one $\mathbf{v}_{i}^{d}$.}

In the standard SFM \cite{helbing2000simulating}, the repulsive forces $\mathbf{f}_{ij}^{p}$ include only a radial component and two terms that account for the friction and compression between the pedestrians. Subsequent works extend the SFM to include a tangential component in the repulsive forces \cite{moussaid2009experimental, guo2014simulation}. In this work, the tangential component defined in \cite{guo2014simulation} is added to the repulsive forces, in order for pedestrians to avoid each other when their paths are crossing.

\new{Let $r_{ij} = r_i + r_j$ be the sum of the radii of agents $i$ and $j$, and $d_{ij} = ||\mathbf{p}_{i} - \mathbf{p}_{j}||$ be the distance between the two agents.}
Let $\mathbf{n}_{ij}$ and  $\mathbf{t}_{ij}$ be right-handed orthogonal unit vectors, such that $\mathbf{n}_{ij}$ is the unit vector of $\mathbf{p}_{i} - \mathbf{p}_{j}$.
Then, the repulsive force $\mathbf{f}_{ij}^{p}$ exerted by the $j$-th pedestrian on the $i$-th one is given by
\begin{equation}
    \mathbf{f}_{ij}^{p} = \left[ A_i e^{(r_{ij} - d_{ij}) / B_i} + k_1 \max\{0, r_{ij} - d_{ij}\} \right] \mathbf{n}_{ij} + \left[ C_i e^{(r_{ij} - d_{ij}) / D_i} + k_2 \max\{0, r_{ij} - d_{ij}\} \Delta v_{ji}^{(t)} \right] \mathbf{t}_{ij}, \label{eq:fij_2}
\end{equation}
where $\Delta v_{ji}^{(t)} = (\mathbf{v}_{j} - \mathbf{v}_{i})^{T}\mathbf{t}_{ij}$ and
$A_i$, $B_i$, $C_i$, $D_i$, $k_1$, $k_2$ are model parameters. 
The two terms on the right hand side of \eqref{eq:fij_2} are, respectively, the radial and tangential components of the repulsive force. 
\new{Each term is the sum of two parts: the first one decreases exponentially with the distance between the agents; the part containing the expression $\max\{0, r_{ij} - d_{ij}\}$ refers to the compression and friction components of the force, respectively, which are active only when $d_{ij} < r_{ij}$, i.e., when the two agents are overlapping. }

\subsection{Headed Social Force Model}
\begin{figure}[t]
  \centering
  \includegraphics[width=\columnwidth]{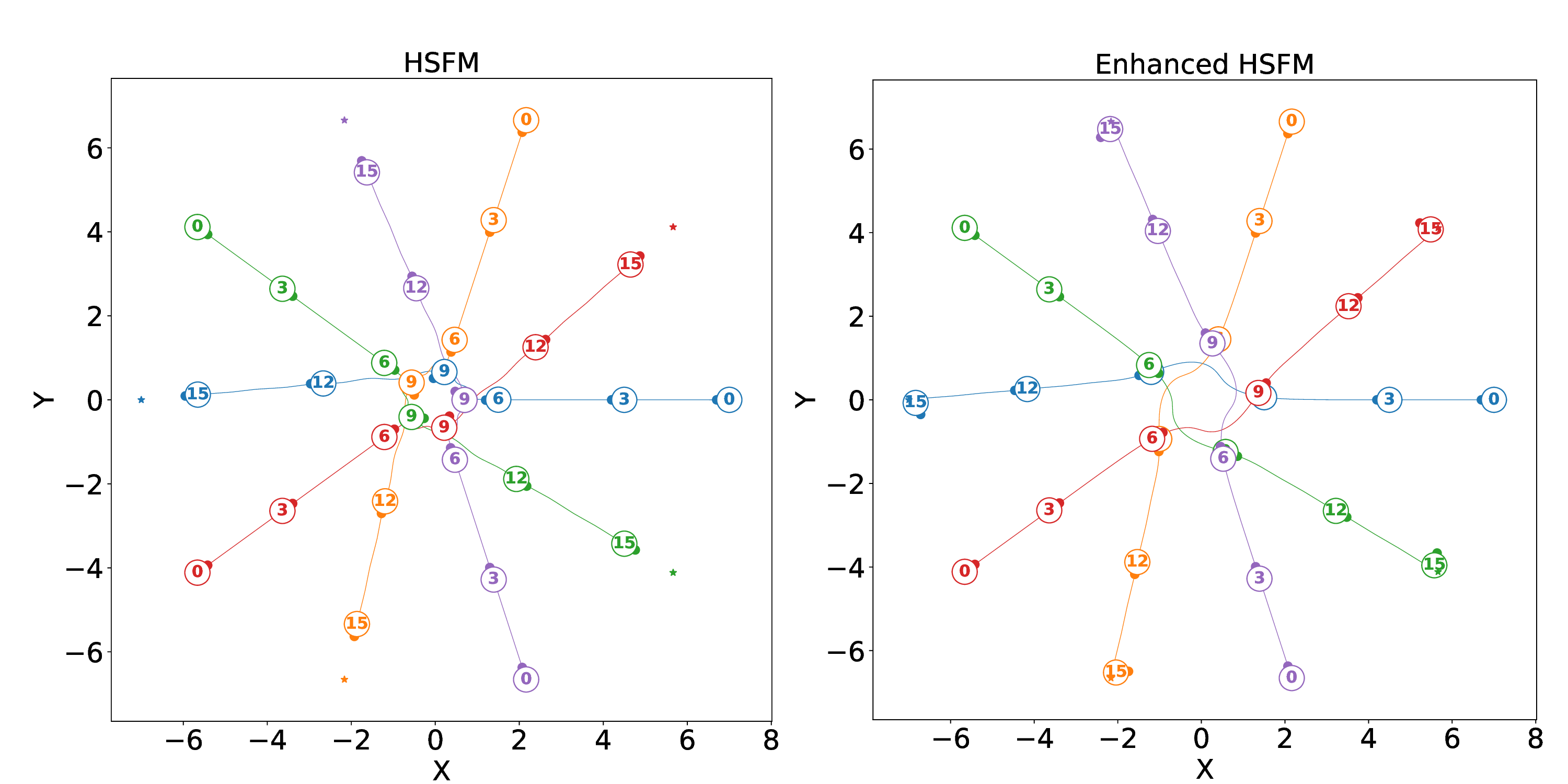}
  \caption{Trajectories of 5 pedestrians (represented as colored circles) following two different versions of the HSFM: \cite{HSFM} (left); this work (right). The configurations of the pedestrians are depicted every three seconds (the time is indicated in the middle of each circle) while the entire path is represented with a segment of the same color of the circle. The star symbols indicate the goals of each pedestrian.}\label{fig:f0}
\end{figure}

To improve the realism of the trajectories generated by the SFM, the Headed Social Force Model (HSFM) explicitly considers also the heading of each pedestrian  \cite{HSFM}. To this purpose, it is convenient to attach a \emph{body frame} to each individual, centered at the pedestrian position and whose x-axis is directed towards the pedestrian heading $\theta_i$. Let us denote  by $\mathbf{v}_i^B = [v_i^f, v_i^o]^T$ the velocity vector expressed in the body frame. Thus, the velocity vector expressed in the global frame can be defined as $\mathbf{v}_i = \mathbf{R}(\theta_i) \mathbf{v}_i^B$, through the rotation matrix
\begin{equation}
    \mathbf{R}(\theta_i) = \left[ \begin{matrix}
    \cos(\theta_i) & -\sin(\theta_i) \\
    \sin(\theta_i) & \cos(\theta_i)
    \end{matrix} \right] = [\mathbf{r}_i^f,\mathbf{r}_i^o],
\end{equation}
\new{where $\mathbf{r}_i^f$ and $\mathbf{r}_i^o$ are the unit vectors identifying the forward and orthogonal direction of motion.} In the HSFM, the state of the $i$-th agent contains their position and orientation, along with the linear and angular velocity, i.e., $\mathbf{x}_i=[\mathbf{p}_i^T, \mathbf{v}_i^T,\theta_i,~\omega_i]^T$. The input is composed by the forces and torques acting on the pedestrian, namely $\mathbf{u}_i=[{u}_{i}^f, {u}_{i}^o, u_i^{\theta}]^T$, where ${u}_{i}^f, {u}_{i}^o$ are the forces acting in the forward and orthogonal direction, respectively, and ${u}_{i}^{\theta}$ is the torque.
Then, the motion dynamics is described by the equations
\begin{gather}
    \dot{\mathbf{p}}_{i} = \mathbf{R}(\theta_i) \mathbf{v}_{i}^B, \\
    \dot{\mathbf{v}}_{i}^B = \frac{1}{m_{i}}\mathbf{u}_{i}^B, \\
    \dot{\theta}_i = \omega_i, \\
    \dot{\omega}_i = \frac{1}{I_i} u_i^{\theta},
\end{gather}
\new{where $\mathbf{u}_{i}^B = [{u}_{i}^f, {u}_{i}^o]^T$ is the force vector in the body frame and $I_i$ denotes the moment of inertia of the $i$-th pedestrian.}

The driving mechanism expresses the way forces and torques are computed, as a function of the system state and the pedestrian goal. The forces acting on the $i$-th agent  are given by
\begin{gather}
    {u}_{i}^f = (\mathbf{f}_{i}^{d} + \sum_{j \neq i} \mathbf{f}_{ij}^{p} )^T \mathbf{r}_i^f, \\
    {u}_{i}^o = k^o (\sum_{j \neq i} \mathbf{f}_{ij}^{p} )^T \mathbf{r}_i^o - k^d v_i^o ,
\end{gather}
where $\mathbf{f}_{i}^{d}$ and $\mathbf{f}_{ij}^{p}$ are computed as in \eqref{eq:fi0}-\eqref{eq:fij_2}. The terms $k^o$ and $k^d$ are constant parameters of the model. The forward component of the global force is essentially the projection of the total force onto the forward direction of the body frame. \new{The orthogonal component of the force involves only the repulsive forces, projected onto the orthogonal direction $\mathbf{r}_i^o$, and a damping term proportional to the sideward velocity $v_i^o$.}

The input torque is computed as
\begin{equation}
    u_i^\theta = -k^\theta (\theta_i - \theta_i^{d}) -k^\omega \omega_i,
\end{equation}
where $\theta_i^{d}$ is the desired orientation that the agent aims to achieves. The parameters $k^\theta$ and $k^\omega$ are designed so that the actual orientation $\theta_i$ effectively tracks the desired one $\theta_i^{d}$.

In the paper introducing HSFM \cite{HSFM}, $\theta_i^{d}$ was selected as the angle of the desired force $\mathbf{f}_{i}^{d}$. In this work, in order to further promote reciprocal avoidance between humans, $\theta_i^{d}$ is chosen as the angle of the total force $\mathbf{f}_i^{tot} = \mathbf{f}_{i}^{d} + \sum_{j \neq i} \mathbf{f}_{ij}^{p}$ acting on the $i$-th pedestrian. Notice that in both cases, $\theta_i^{d}$ turns out to be a function of $\mathbf{p}_i^{goal}$, through \eqref{eq:fi0} and \eqref{eq:desvel}.

A comparison between the two versions of the HSFM is shown in Figure~\ref{fig:f0}. In this example, five pedestrians cross the environment, starting from different positions along a circumference. In the left figure, the agents move according to the standard HSFM, while on the right they follow the modified version adopted in this work. In the former case, the pedestrians almost collide when they reach the center of the scene. Conversely, in the latter they manage to stay away from each other, because their orientation tends to track the direction of the global force, which includes also the repulsive terms generated by the other agents.

Note that the aforementioned models are described as continuous time dynamical systems but, in order to simulate human motion, the models are discretized with an appropriate sampling time $\Delta t$.


\section{Robot navigation policies}\label{nav_alg}
In this section, the navigation policies used in this work to drive the robot in a human populated environment are introduced. Among the variety of navigation algorithms available in the literature, here we focus on policies based on reinforcement learning (RL). They rely on the definition of an optimal state-value function $V^*$ (\textit{value function} in the following) which represents the expected cumulative discounted reward (\textit{return} in the following) from a given system state when following an optimal policy $\pi^*$. The reward function $R$ is designed so as to promote the approach of the robot to the goal while penalizing collisions. The value function $V^*$ is approximated by means of a neural network (\textit{value network} in the following), whose weights are learned through experience sets during the training process. The trained value network is then used in the navigation process to compute the "best" robot action given the current system state. This is done by searching for the robot action which maximizes the sum of the immediate reward and the expected return from the state visited by the system at the next time instant.

In this context, the role of the human motion model used to simulate humans in the experience sets becomes crucial. Different models lead to different human behaviors which in turn result in different estimates of the value network. Thus, the robot will end up having an intrinsic understanding of the crowd motion in the operating environment.

Let us introduce a common framework for the policies considered in this work. To this purpose, the RL formulation of a navigation problem involving one robot and one human agent $i$ is summarized, following the notation proposed in \cite{chen2017decentralized}.
The robot is represented by a disc of radius $r_r$, centered in the position vector $\mathbf{p}_r$. The robot velocity $\mathbf{v}_r$ is the action $\mathbf{u}_r$ to be determined by the navigation policy. The holonomic motion model
\begin{equation}
    \dot{\mathbf{p}}_r = \mathbf{u}_r
\end{equation}
is assumed for the robot. The action space of the robot is the set $U = \{ \mathbf{u}_r \, | \, ||\mathbf{u}_r|| < v_r^{max} \}$, where $v_r^{max}$ is the maximum robot speed.

From a RL perspective, the state of the system contains all the relevant information on the robot and the human. The robot state $\mathbf{s}$ and the human state $\mathbf{\tilde s}$ contain the position and velocity of the corresponding agent, the position they aim to reach, as well as their radius and maximum speed, i.e. $\mathbf{s} = (\mathbf{p}_r, \mathbf{v}_r, \mathbf{p}_r^{goal}, r_r, v_r^{max})$ and $\mathbf{\tilde s} = (\mathbf{x}_i, \mathbf{p}_i^{goal}, r_i, v_i^{max})$.

The human state $\mathbf{x}_i$ depends on the considered human motion model, as illustrated in the previous section. For instance, in the HSFM model $\mathbf{x}_i$ includes the orientation and angular velocity of the pedestrians as well.
It is assumed that when deciding its action, the robot has access to some pieces of information about the human. These consist of the observable part of the pedestrian state, i.e.,  $\mathbf{\tilde s^{o}} = (\mathbf{x}_i, r_i)$. Hence, the action of the robot is selected according to a policy $\pi$ which is a function of the joint state $\mathbf{s}^{jn} = (\mathbf{s},  \mathbf{\tilde s}^{o})$, i.e. $\mathbf{u}_r = \pi(\mathbf{s}^{jn})$.

\begin{figure}[t]
    \centering
    \includegraphics[width=0.8\columnwidth]{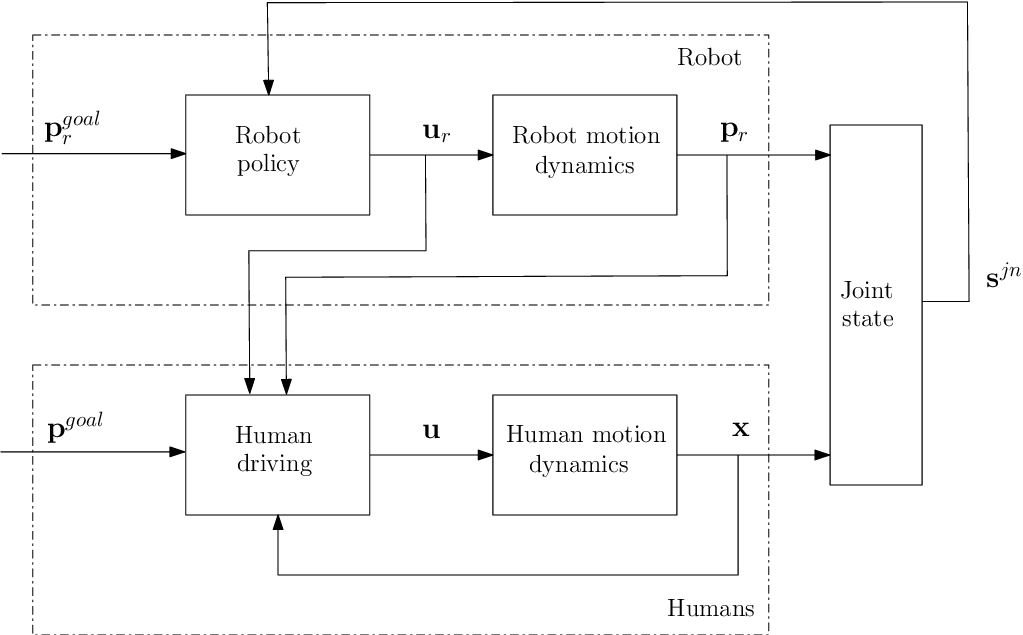}
    \caption{Diagram of the interaction between robot and human motion models. The position $\mathbf{p}_r$ and velocity $\mathbf{u}_r$ of the robot are passed to humans only if the former is visible to the latter.}
    \label{fig:simulation_flow}
\end{figure}

The reward function is designed in order to award the robot for reaching the goal and to penalize it if it gets too close to the human or, even worse, if it collides with them. It is defined as
\begin{equation}
    R(\mathbf{s}^{jn},\mathbf{u}_r) = 
        \begin{cases}
          -0.25 & \text{if} \qquad d_{min} < 0\\
          -\frac{(d_s - d_{min})\Delta t}{2} & \text{if} \qquad 0 \leq d_{min} < d_s\\
          1 & \text{if} \qquad ||\mathbf{p}_r - \mathbf{p}_r^{goal} || < r_r\\
          0 & \text{otherwise}
        \end{cases}\,,
        \label{eq:reward}
\end{equation}
where $d_s$ is a separation distance the robot should keep from the pedestrian and
$d_{min}$ is the minimum distance between the robot and the human over the next time interval $\Delta t$, assuming a constant human velocity. Notice that $d_{min}$ is a function of the current position and velocity of both the robot and the human. When the robot uses an optimal navigation policy $\pi^*$, the value function from state $\mathbf{s}^{jn}_t$ at time $t$ is defined as (eq. (3.11),(3.12) in \cite{sutton2018reinforcement})
\begin{equation}
    V^*(\mathbf{s}^{jn}_t) = \sum_{\tau=t+1}^T \bar \gamma^{\tau-t-1} R(\mathbf{s}^{jn}_\tau, \pi^*(\mathbf{s}^{jn}_\tau))
        \label{eq:valuef}
\end{equation}
where the discount factor $\bar \gamma$ depends on the sampling time $\Delta t$ and maximum robot speed $v_r^{max}$ (e.g., $\bar{\gamma} = \gamma^{\Delta t \cdot v_r^{max}}$, with $0<\gamma<1$) and $T$ represents the length of the episode. 
From the value function $V^*$, an optimal policy $\pi^*$ can be derived as \begin{equation}\label{eq:cadrl_policy}
    \pi^{*}(\mathbf{s}^{jn}_t) = \arg \max_{\mathbf{u}_r \in U} \left( R(\mathbf{s}^{jn}_t,\mathbf{u}_r) + \bar{\gamma}  \int_{\mathbf{s}^{jn}_{t+1}} P(\mathbf{s}^{jn}_{t+1} \, | \, \mathbf{s}^{jn}_t, \mathbf{u}_r) V^{*}(\mathbf{s}^{jn}_{t+1}) \, d\mathbf{s}^{jn}_{t+1}  \right),
\end{equation}
where $P(\mathbf{s}^{jn}_{t+1} \, | \, \mathbf{s}^{jn}_t, \mathbf{u}_r)$ are the state transition probabilities of the system (which depend on the human motion model). Hence, the optimal robot action $\mathbf{u}_r^*$ at time $t$ is computed from the current state $\mathbf{s}^{jn}_t$ as $\mathbf{u}_r^* = \pi^*(\mathbf{s}^{jn}_t)$.

Figure~\ref{fig:simulation_flow} illustrates the interaction between robot and human motion models. Notice that, when the robot is visible to humans, the human driving mechanism receives in input the position $\mathbf{p}_r$ and velocity $\mathbf{u}_r$ of the robot. This allows the pedestrians to consider the robot just as one of the agents in the scene.

In an RL setting, the true value function $V^*(\cdot)$ is typically approximated by using a value network $V(\cdot;\mathbf{W})$, whose weights $\mathbf{W}$ are estimated by a training procedure.  The specific RL-based navigation policies \cite{chen2017decentralized,everett2018motion,SARL} considered in this article differ for the architecture of the value network, as well as for the activating functions that are used. Their main features are summarized in the following.

\subsection{Collision Avoidance Deep Reinforcement Learning}\label{cadrl}

In Collision Avoidance Deep Reinforcement Learning (CADRL) \cite{chen2017decentralized}, a three-layer, fully connected value network is used to estimate the value of the joint states. ReLU activation functions are adopted. The weights are optimized during a two-step training process. First, an imitation learning phase is performed using a set of robot and human trajectories generated according to the ORCA model. Then, the value network is refined through RL, using an $\varepsilon$-greedy policy with $\varepsilon$ decreasing during the learning process. 
CADRL can be extended to the case involving $n$ pedestrians, by defining a joint state for each pair of robot and human $\mathbf{s}_{i}^{jn}=(\mathbf{s},  \mathbf{\tilde s}_i^{o} )$, $i=1,\dots,n$, and then choosing the action that maximizes the minimum state value among all the joint states.

\subsection{Long Short-Term Memory Reinforcement Learning}

An approach more suitable to deal with multiple humans, based on Long Short-Term Memory reinforcement learning (LSTM-RL), is proposed in \cite{everett2018motion}.
A recurrent neural network architecture, exploiting a sequence of LSTM cells, is devised in order to handle a variable number of humans. The $i$-th LSTM cell is fed with the observable state $\mathbf{\tilde s}_i^{o}$ of the $i$-pedestrian, together with the output of cell $i-1$. The human agents are sorted by decreasing distance to the robot, so that the closest human to the robot has the biggest effect on the final output. The final LSTM cell outputs a fixed-length hidden state which encodes the relevant information about all the pedestrians in the scene. The hidden state is concatenated with the robot state $\mathbf{s}$ and then given in input to a two-layer, fully connected feedforward network with ReLU activation functions. The overall value network takes as input the joint state $\mathbf{s}^{jn}=(\mathbf{s}, \mathbf{\tilde s}_1^{o},\dots, \mathbf{\tilde s}_n^{o})$ and outputs the value of the joint state, as well as the optimal action corresponding to that state.\footnote{In LSTM-RL, the action space is discrete and the learned policy is stochastic. As a consequence, the output of the network is the probability distribution of the possible actions for the considered joint state.} Hence, LSTM-RL is an actor-critic method which uses a single
network to approximate both the actor and the critic. The overall value network is trained by means of an initial imitation learning (using existing CADRL trajectories), followed by a policy optimization phase.

\subsection{Social Attentive Reinforcement Learning}

Another RL-based approach that takes into account multi-agent interactions is Social Attentive Reinforcement Learning (SARL) \cite{SARL}, which implements a socially attentive network encoding the impact of the entire crowd into the value network. 
The main architecture is composed of three modules. The interaction module models pairwise human-robot interaction. To this purpose, local maps are built in order to take into account the effect of multiple pedestrians close to each other.
The pooling module provides a way to have a fixed-length representation of the crowd, regardless of the number of humans present in the environment. In order to identify the local importance of each robot neighbor, an attention score is computed for each human as the output of a multi-layer network with ReLU activating functions. Then, the attention scores are used to combine together the pairwise human-robot interactions.
Finally, the planning module estimates the value function by means of a multi-layer neural network (with ReLU activating functions) which takes as input the robot state $\mathbf{s}$ and the crowd representation returned by the pooling module.
Also in this case, the weights of all the networks are trained with imitation learning and $\varepsilon$-greedy policy optimization, with decreasing $\varepsilon$.


\begin{table}
    \begin{center}
    \begin{tabular}{ |p{4cm}|p{10cm}| }
     \hline
     \multicolumn{2}{|c|}{\textbf{Circular Crossing Scenario Card}} \\
     \hline
     \hline
     \multicolumn{2}{|l|}{\textbf{Scenario Metadata}} \\
     \hline
     Scenario Name & Circular crossing \\
     \hline
     Scenario Description & Humans and robot are randomly placed on the border of a circle with a predefined radius. Each agent must reach the point diametrically opposite to its initial configuration on the circle.\\
     \hline
     Scientific Purpose & Crowded scenario applicable indoors and outdoors in wide open spaces.\\
     \hline
     \multicolumn{2}{|l|}{\textbf{Scenario Definition}} \\
     \hline
     Geometric Layout & A circle of 7 meter radius. \\
     \hline
     Intended Robot Task & Reach the target within 50 seconds without collisions. \\
     \hline
     Intended Human Behavior & Navigate back and forth from their initial configuration to their target according to the chosen human motion model. \\
     \hline
     \multicolumn{2}{|l|}{\textbf{Scenario Usage Guide}} \\
     \hline
     Success Metrics & Success rate. \\
     \hline
     Ideal Outcome & Robot reaches the target in a socially acceptable manner.\\
     \hline
     Failure Modes & Robot collides with a human or fails to reach the target within the time limit. \\
     \hline
    \end{tabular}
    \end{center}
    \caption{Information regarding the circular crossing scenario.}\label{tab:cc_scenario}
\end{table}

\begin{figure}[t]
    \centering
    \includegraphics[width=0.9\columnwidth]{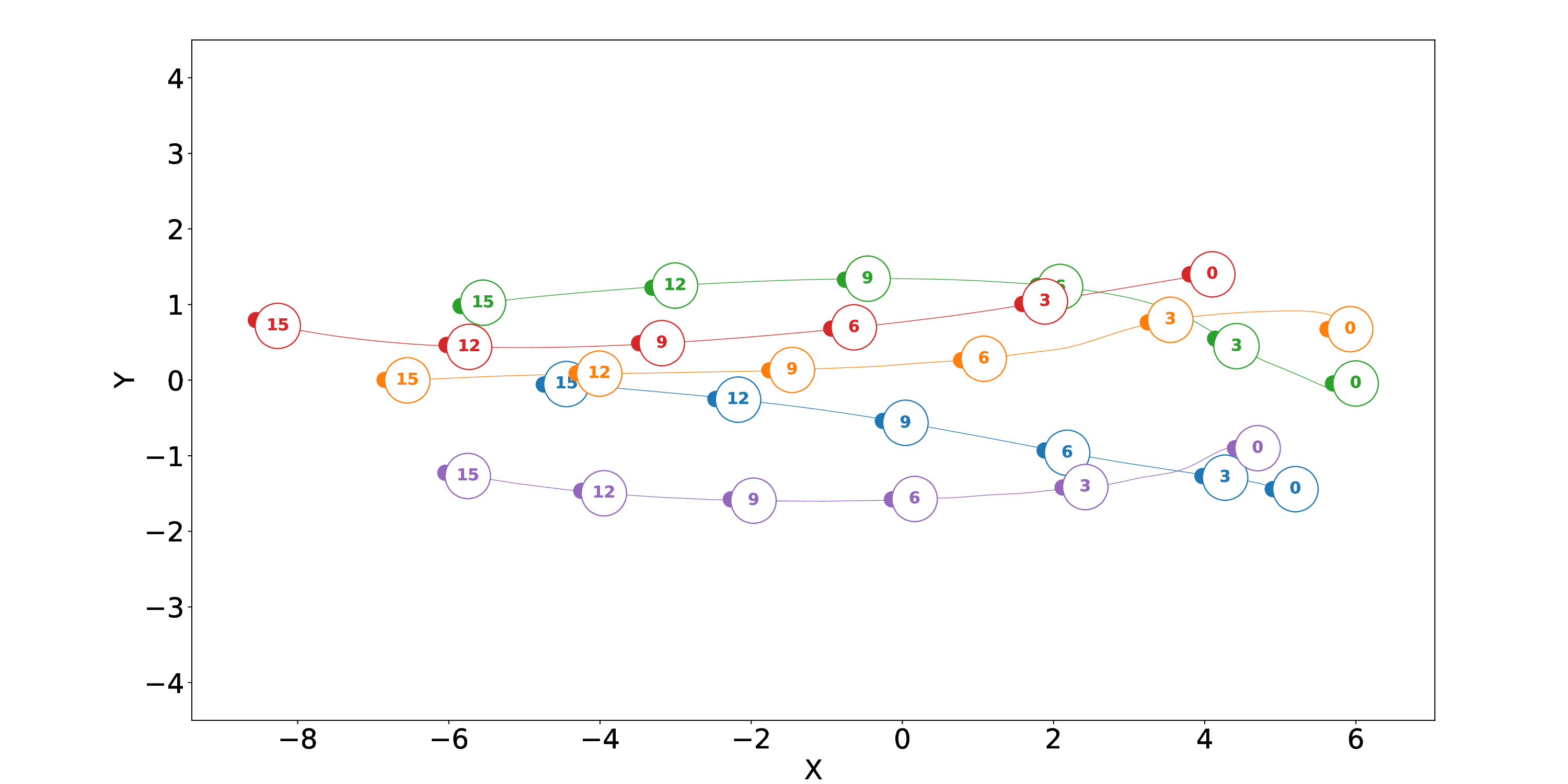}
    \caption{Trajectories of 5 humans moving in the parallel traffic scenario driven by the HSFM. The colored lines represent the trajectory of each human. The position and orientation of each agent are shown every 3 seconds.}\label{fig:parallel_traffic}
\end{figure}

\section{Simulation scenarios}\label{scenarios}
In this section, the scenarios employed during the training and testing phase of the robot navigation policies are described. Following the protocol defined in \cite{francis2023principles}, the corresponding scenario cards are provided.

The first scenario considered in this work is the Circular Crossing (CC), similar to the one addressed in \cite{mavrogiannis2023core}, in which a robot and a predefined number of humans are initially placed along the boundary of a circle at random. The task of each agent (humans and robot) is to reach the point of the circumference that is diametrically opposite to their initial configuration. An example of humans moving according to the HSFM in the CC scenario can be seen in Figure~\ref{fig:f0}. Such a  scenario is often used for social navigation studies because it forces the robot to deal with a high-density crowd when every agent approaches the center of the scene. The circular crossing scenario card is reported in Table \ref{tab:cc_scenario}.

The second scenario analyzed, known as Parallel Traffic (PT) \cite{francis2023principles}, simulates a situation in which there is a dense crowd of pedestrians continuously flowing in one direction, while the robot has to move in the opposite direction of the crowd to reach its target. An example of 5 humans moving in the parallel traffic scenario driven by the HSFM is reported in Figure~\ref{fig:parallel_traffic}. In this case, the robot starts from $\mathbf{p}_{0} = [-6,0]$ to reach its goal at $\mathbf{p}^{goal} = [6,0]$, having to move against the crowd. For this scenario, the aim is to test the ability of the robot to avoid incoming agents in a socially acceptable way.

The parallel traffic scenario card is shown in Table \ref{tab:pt_scenario}.
\begin{table}[h]
    \begin{center}
    \begin{tabular}{ |p{4cm}|p{10cm}| }
     \hline
     \multicolumn{2}{|c|}{\textbf{Parallel Traffic Scenario Card}} \\
     \hline
     \hline
     \multicolumn{2}{|l|}{\textbf{Scenario Metadata}} \\
     \hline
     Scenario Name & Parallel traffic. \\
     \hline
     Scenario Description & A crowd of humans flows in one direction in a rectangle-shaped space while the robot drives in the opposite direction to reach its target. \\
     \hline
     Scientific Purpose & Crowded scenario applicable indoors and outdoors in narrow spaces or lanes.\\
     \hline
     \multicolumn{2}{|l|}{\textbf{Scenario Definition}} \\
     \hline
     Geometric Layout & A rectangle of base 14 meters and height 3 meters. \\
     \hline
     Intended Robot Task & Reach the target within 50 seconds without collisions. \\
     \hline
     Intended Human Behavior & Continuously flow in the given space and in the same direction according to the chosen human motion model. Whenever a human reaches the end of the space, another one enters on the other side at the same height. \\
     \hline
     \multicolumn{2}{|l|}{\textbf{Scenario Usage Guide}} \\
     \hline
     Success Metrics & Success rate. \\
     \hline
     Ideal Outcome & Robot reaches the target in a socially acceptable manner.\\
     \hline
     Failure Modes & Robot collides with a human or fails to reach the target within the time limit. \\
     \hline
    \end{tabular}
    \end{center}
\caption{Information regarding the parallel traffic scenario.}\label{tab:pt_scenario}
\end{table}


\section{Evaluation metrics}\label{metrics}
To evaluate the performance of each navigation policy, two different types of metrics are considered: social and navigation metrics \cite{francis2023principles}. Navigation metrics  evaluate the efficiency and effectiveness of the navigation policy from the perspective of the navigation task. Such metrics include, among others, success rate, number of collisions and time to goal. On the other hand, social metrics focus on the social aspects of the navigation experience, such as the average distance to humans within a given trajectory or the average robot jerk, which may be disturbing for humans.

The navigation metrics and parameters used in this study are described below.
\begin{itemize}
    \item Success rate: the ratio between the number of episodes in which the robot reaches its goal without colliding with any human, within the prescribed time limit, and the total number of episodes.
    \item Time to goal: an episodic metric indicating the time between the beginning of an episode and the time when the robot reaches its target. Clearly, this metric is computed only if an episode is successful.
    \item Average speed: the average norm of the robot velocity in a successful episode.
    \item Success weighted by path length (SPL) \cite{anderson2018evaluation}: computed over a given number of episodes, it is the success rate penalized by the normalized inverse length of all the successful trajectories. It is defined as
    \begin{equation}
        SPL = \frac{1}{N}\sum^{N}_{i=1} S_{i} \frac{l_i}{p_i},
    \end{equation}
    where $S_i$ is a binary variable equal to 1 if the episode was successful and to 0 otherwise; $N$ is the total number of episodes; $l_i$ the shortest path length between the robot initial position and its target; $p_i$ the length of the actual path taken by the robot.
\end{itemize}
The following social metrics are employed in this study.
\begin{itemize}
    \item Space compliance: an episodic metric representing the percentage of time in which the robot minimum distance to humans is over 50 centimeters (also known as personal space compliance) in a successful episode.
    \item Average acceleration: the average norm of the robot acceleration in a successful episode.
    \item Average jerk: the average norm of the derivative of the acceleration in a successful episode.
    \item Average minimum distance to humans: the average minimum distance between the robot and humans within a successful episode.
\end{itemize}

Table \ref{tab:metrics} summarizes all the considered metrics along with their main features. \new{The last column indicates whether the preferred values for the considered metric should be high or low, given the social or navigational focus of the metric itself. For example, a low time to goal is preferable for navigation efficiency, while a high space compliance is required for social acceptability.}

\begin{table}
    \begin{center}
    \begin{tabular}{ |c|c|c|c| }
        \hline
        \textbf{Metric} & \textbf{Episodic} & \textbf{Social/Navigation} & \textbf{Preference(High/Low)} \\
        \hline
        Success rate & No & N & H \\
        \hline
        Time to goal & Yes & N & L \\
        \hline
        Average speed & Yes & N & H \\
        \hline
        SPL & No & N & H \\
        \hline
        Space compliance & Yes & S & H \\
        \hline
        Average acceleration & Yes & S & L \\
        \hline
        Average jerk & Yes & S & L \\
        \hline
        Average min. dist. to humans & Yes & S & H \\
        \hline
    \end{tabular}
    \end{center}
\caption{Summary table of the evaluation metrics employed in this study. The last column indicates the preference for the values of each metric.}\label{tab:metrics}
\end{table}

\newpage
\section{Numerical results}\label{num_results}
In this section, the training and testing procedures carried out are first introduced. Then, the test results are analyzed and thoroughly discussed.

\subsection{Simulation software}
To train and test the robot navigation policy, a dedicated software (Social-Navigation-PyEnvs\footnote{Available at \url{https://github.com/TommasoVandermeer/Social-Navigation-PyEnvs}}) was developed in Python, starting from the CrowdNav module\footnote{Available at \url{https://github.com/vita-epfl/CrowdNav}.} developed in \cite{SARL} and the Python-RVO2 module\footnote{Available at \url{https://github.com/sybrenstuvel/Python-RVO2}}. This software combines the implementations of the human motion models and the robot navigation algorithms provided by CrowdNav, to yield a module that includes a 2D GUI developed with the Pygame library\footnote{Available at \url{https://www.pygame.org/}}. As suggested in \cite{everett2018motion}, the sampling time $\Delta t$ for simulations has been carefully tuned. In particular, for exploration purposes, the sampling time needs to be large so that the next joint states $\mathbf{s}^{jn}_{t+1}$ are sufficiently spread. However, if $\Delta t$ is set too large there is an increase in the collisions with humans due to the fact that decisions are not taken frequently enough. Moreover, a too large sampling time may cause numerical instability when integrating human dynamics. For these reasons, two different sampling times are used for humans and robot. For the robot, the control action is applied every $\Delta t_{r} = 0.25s$, while for humans a sampling time $\Delta t = 0.01s$ is employed, ensuring accurate numerical integration of the human dynamics. Each RL policy was trained for 10.000 training episodes and took about 23 hours on average on an i9-10980 CPU @ 3.00Ghz (note that the higher frequency used for human dynamics makes the training procedure longer).

The policies trained with HSFM-driven humans include the orientation $\theta_i$ and the angular velocity $\omega_i$ in the state $\mathbf{x}_i$. In order to use them also in environments where humans move according to models in which $\theta_i$ and $\omega_i$ are not defined, the weights of the value network $V(\cdot,\mathbf{W})$ relative to $\theta_i$ and $\omega_i$ are set to zero in the implementation of all the robot navigation policies. Moreover, for SARL policies, the usage of local maps is neglected as they perform similarly to SARL policies that instead make use of them (according to what observed in \cite{SARL}). 

\begin{figure}[t]
    \centering
    \includegraphics[width=120mm]{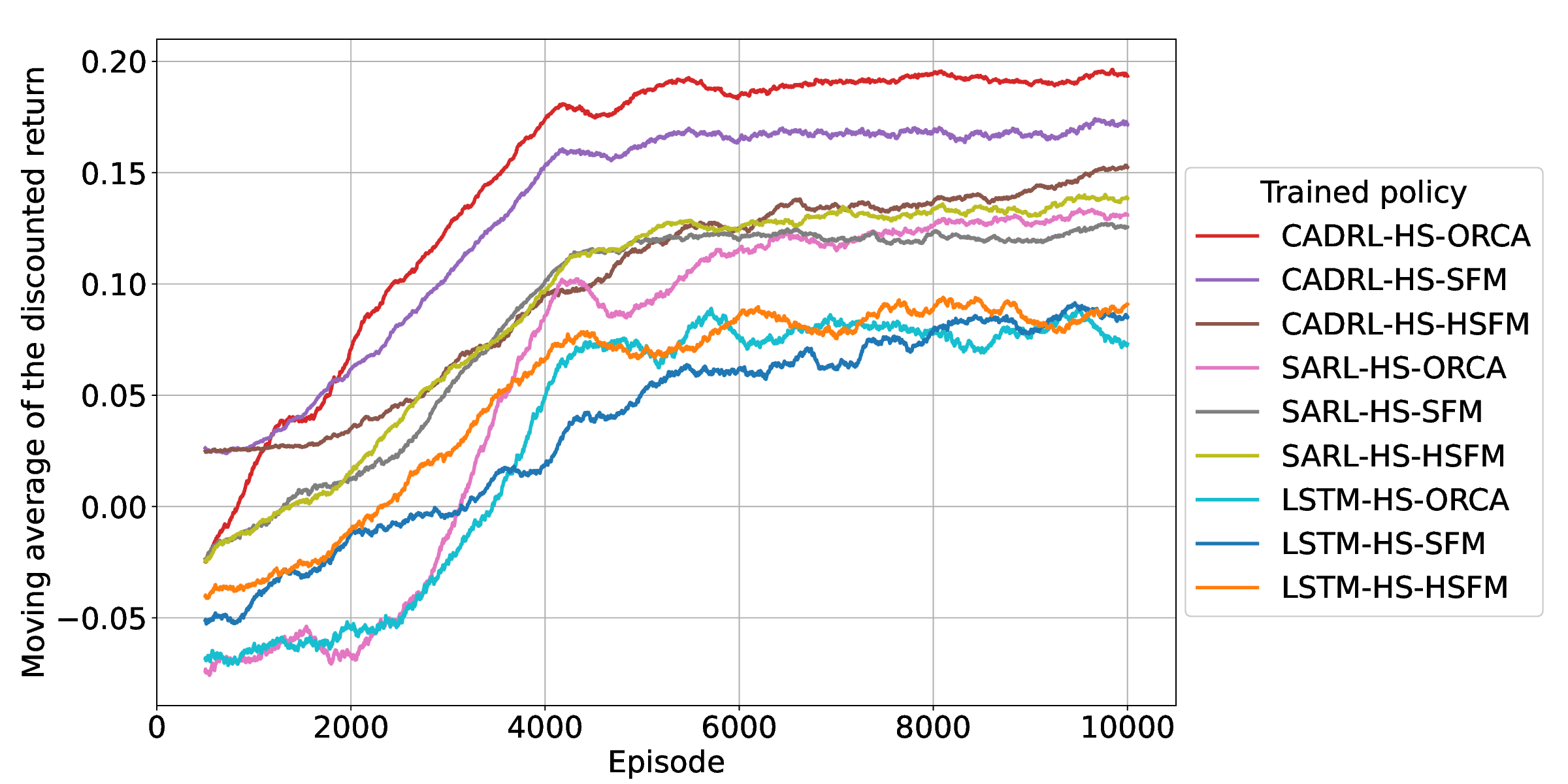}
    \caption{Moving average over 500 episodes of the discounted return (with discount factor $\gamma = 0.9$)  of the robot navigation policies during training in the hybrid scenario.}
    \label{fig:hs_training}
\end{figure}

\subsection{Training and testing settings} 
Each RL policy (CADRL, LSTM-RL, SARL) has been trained in 9 different \textit{training settings}, \new{where each setting is a combination of one human motion model (ORCA, SFM or HSFM, hereafter collectively denoted as \HMM), and one scenario.} Besides the CC and PT scenarios introduced in Section~\ref{scenarios}, an additional Hybrid Scenario (HS) has been used for training, in which the scenario of each episode is chosen at random (with equal probability) between CC and PT. Hence, three possible scenarios are considered for training. As in \cite{chen2017decentralized}, all CADRL policies have been trained in an environment with one human (because CADRL uses a two-agent value network even in the multi-agent case) while SARL and LSTM-RL policies have been trained in environments with 5 humans. In all cases, the robot was not visible to the humans. This prevents humans from taking care of avoiding collisions and makes the training more challenging. Following the approach in \cite{SARL}, the value networks for all the robot policies have been initialized with imitation learning using a set of 3000 demonstrator episodes in which the robot moves according to the same motion model employed for humans. The outcome of the training phase are 27 trained policies in total (3 policies $\times$ 9 training settings). As an example, a possible combination is a trained SARL policy in the CC scenario where humans move according to the ORCA model.

To assess the effectiveness of the training procedure, Figure~\ref{fig:hs_training} reports the moving average over a 500 episodes window of the discounted episodic return of each policy trained in the hybrid scenario. For all the policies the return settles after approximately 6000 training episodes, suggesting that the training process has been completed. Note that the return for CADRL policies is the higher, most likely due to the fact that for these policies there is only one human in the training environment.

Each trained policy has been evaluated in 6 different \textit{testing settings} and with 5 different numbers of humans (5, 10, 15, 20, 25), for 100 trials each. A radius $r_r = r_i = 0.3 m$ and maximum velocity $v_r^{max} = v_i^{max} = 1.0 m/s$ is used for both the robot and humans and a discomfort distance $d_s = 0.2 m$ is employed for the reward function \eqref{eq:reward}. Every testing setting is the result of the combination of one of the two scenarios (CC or PT) and a human motion model (ORCA, SFM or HSFM). Thus, an example of testing setting is the PT scenario with humans driven by the SFM model. In contrast to the training setting, during the testing phase the robot is visible to humans, making the testing environments more realistic.

\subsection{Comparing policies}
\begin{figure}[t]
    \centering
    \includegraphics[width=0.8\columnwidth]{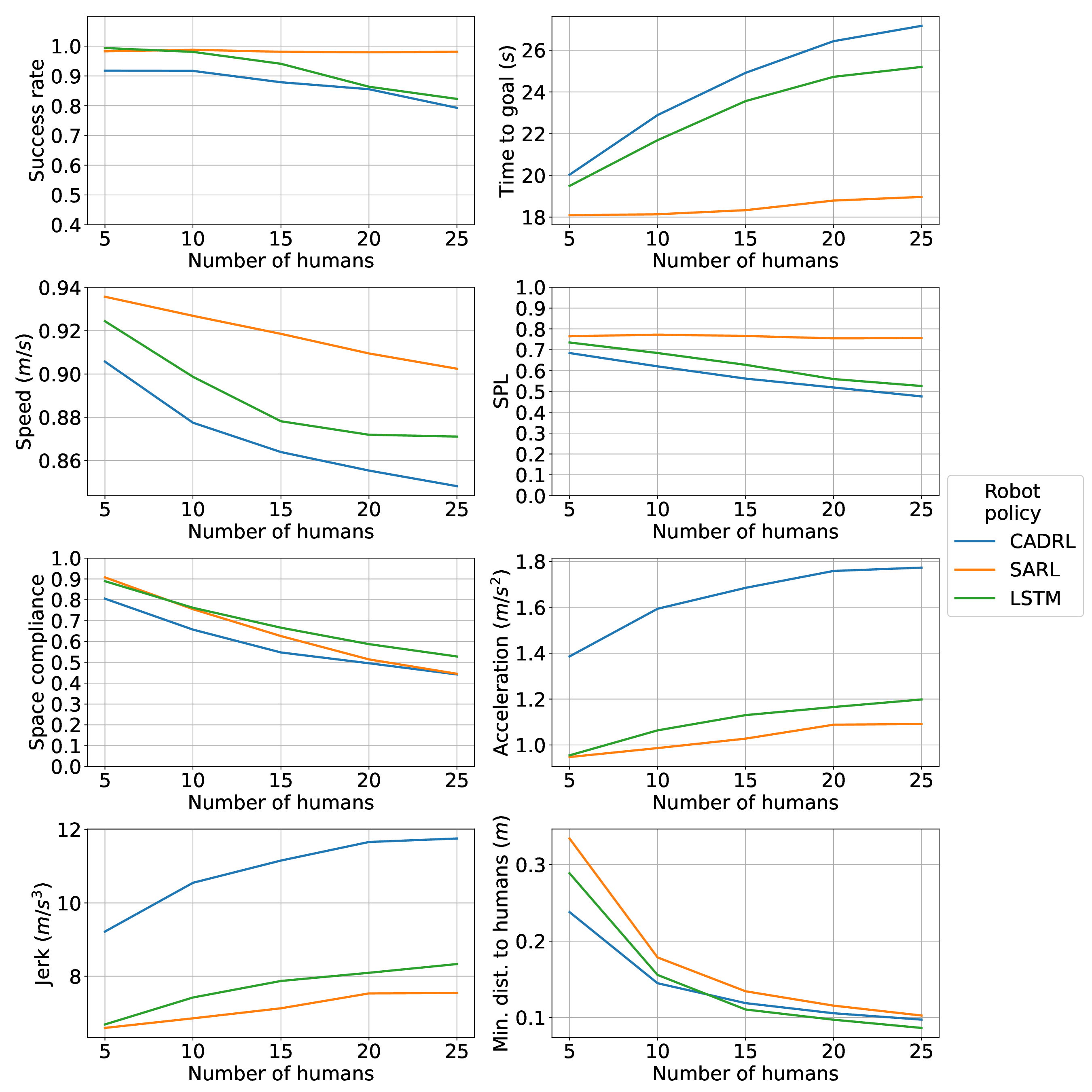}
    \caption{Navigation and social metrics against the number of humans populating the testing environment. Each curve represents the average metric for a robot navigation policy (CADRL, LSTM-RL, SARL) over 5400 episodes (9 training settings $\times$ 6 testing settings $\times$ 100 trials).}
    \label{fig:policies_metrics}
\end{figure}
\begin{figure}[t]
    \centering
    \includegraphics[width=0.8\columnwidth]{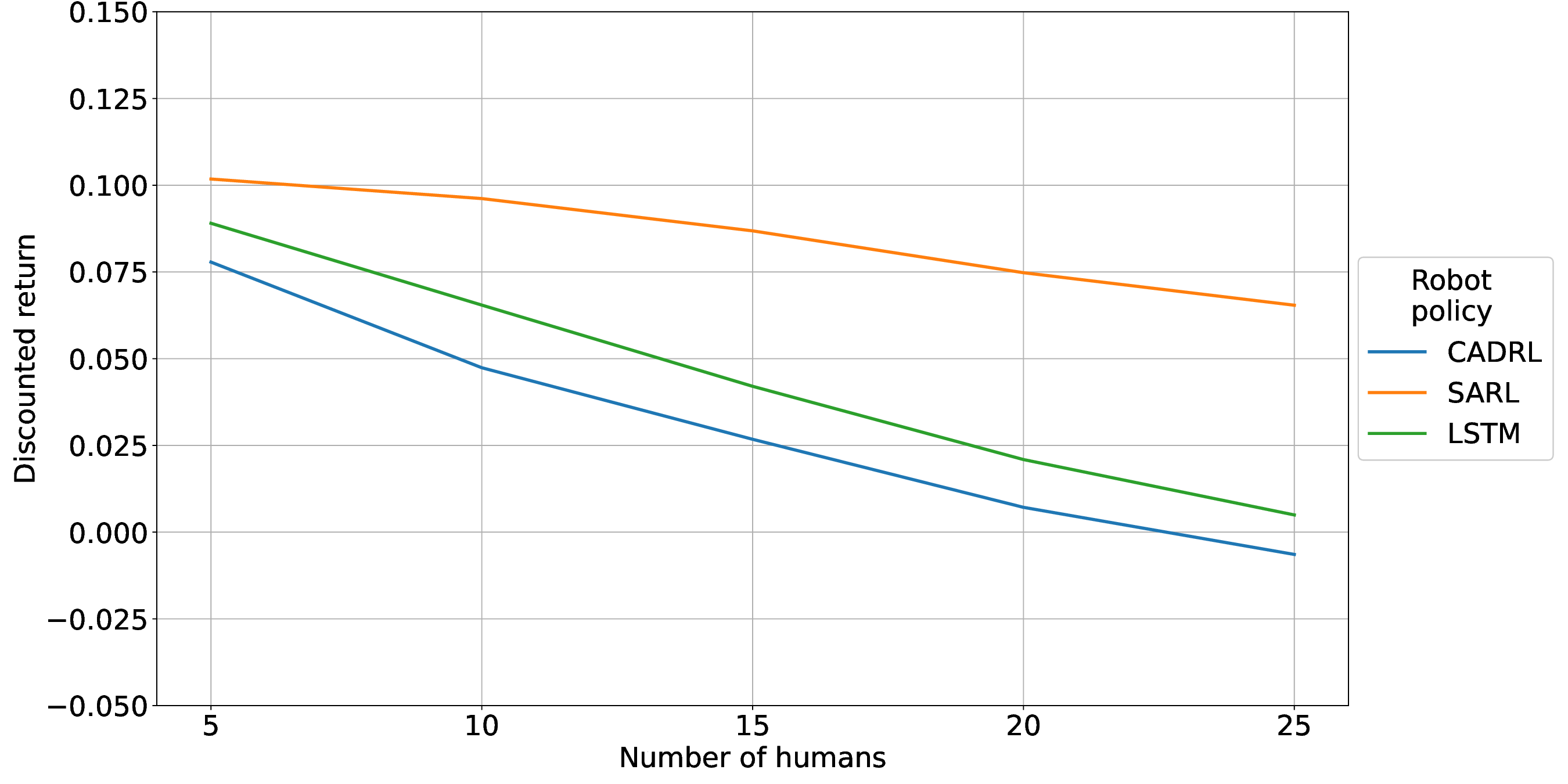}
    \caption{Discounted return (with discount factor $\gamma = 0.9$) against the number of humans populating the testing environment. Each curve is averaged over 5400 episodes (as in Figure~\ref{fig:policies_metrics}).}
    \label{fig:policies_return}
\end{figure}

As a first study, the performance of the three different robot navigation policies has been compared. The navigation and social metrics introduced in Section \ref{metrics} have been computed for each navigation policy (CADRL, LSTM-RL, SARL). The performance has been averaged over the 9 training settings (3 scenarios $\times$ 3 human motion models) and 6 testing settings (2 scenarios $\times$ 3 human motion models). Figure~\ref{fig:policies_metrics} reports the resulting average metrics, evaluated against the number of humans populating the testing environment, to highlight the specific features of the navigation policies. Although all policies are generally successful in navigating the robot through the crowded environment, it can be observed that SARL policies present the highest success rate and they are much less sensitive to an increase in the crowd density with respect to CADRL and LSTM-RL policies. It is worth mentioning that all episode failures are due to exceeding the maximum time limit, not to collisions. In addition, SARL features the lowest mean time to goal while maintaining a space compliance comparable to that of the other policies. The number of humans does not affect the performance of SARL in terms of SPL, while it does for the other policies. Additionally, looking at the social metrics, SARL policies show the lowest acceleration and jerk norms, while maintaining the highest minimum distance to humans. Overall, SARL policies seem to achieve the best compromise between navigation efficiency and social requirements.
In Figure~\ref{fig:policies_return} the average discounted return for the robot navigation policies during tests is reported. As in Figure~\ref{fig:policies_metrics}, the return has been averaged over 9 training settings and 6 testing settings. Once again, SARL policies are the ones yielding the best performance and are less sensitive to an increase in the number of humans populating the environment. Since the success rate for SARL is close to 100\% (as shown in Figure~\ref{fig:policies_metrics}), the discounted return can be interpreted as a metric that accounts for the robot velocity to reach the target (because the reward for reaching the target is discounted by the time required to obtain it) and its ability to maintain a certain distance from humans. Thus, this further confirms the good trade off between navigation performance and social compliance.
\begin{figure}[t]
    \centering
    \includegraphics[width=0.8\columnwidth]{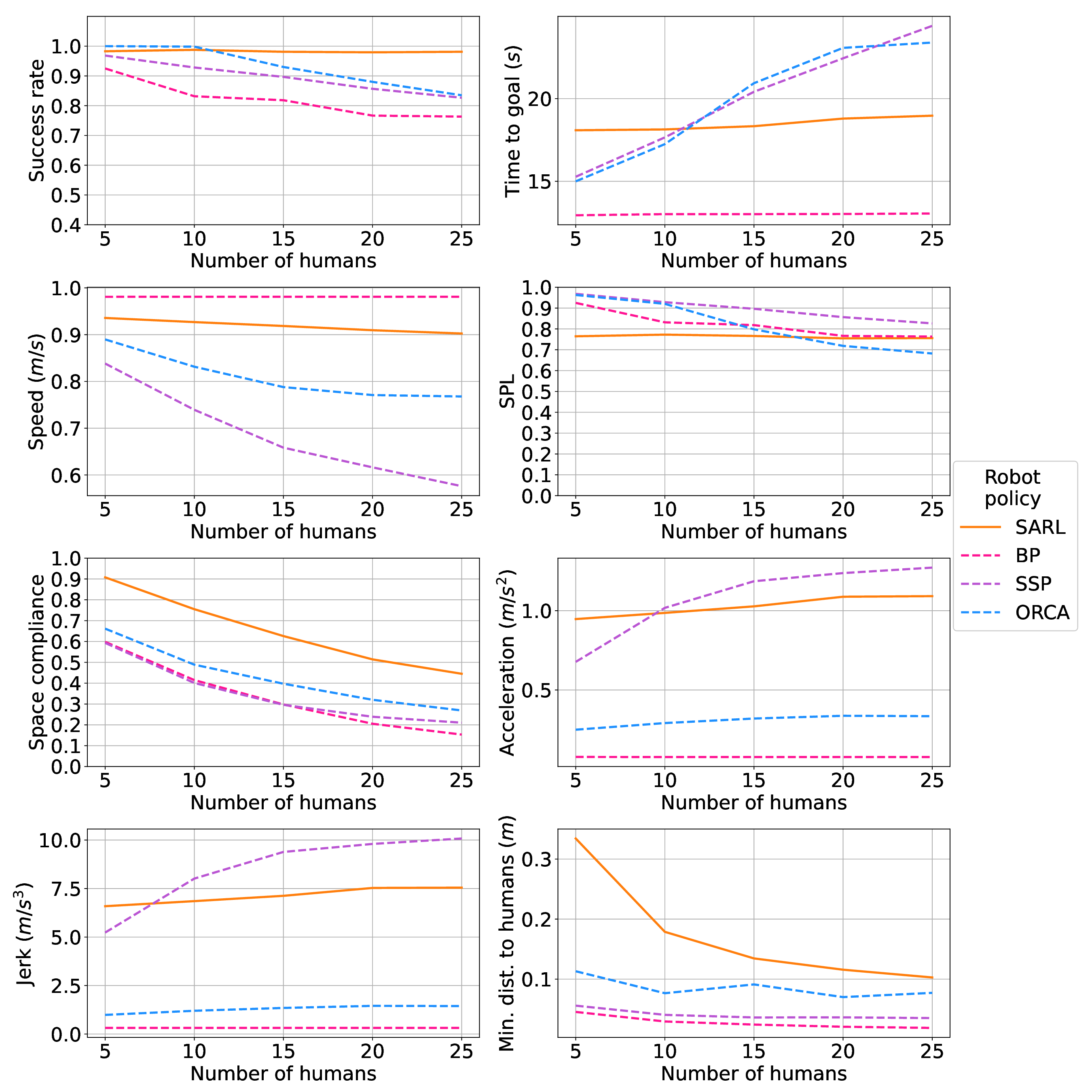}
    \caption{Navigation and social metrics for SARL and baseline policies (BP, SSP, ORCA) against the number of humans populating the testing environment. Each baseline curve is averaged over 600 episodes (6 testing scenarios $\times$ 100 trials).}
    \label{fig:baselines_against_sarl}
\end{figure}
\begin{figure}[t]
    \centering
    \includegraphics[width=0.8\columnwidth]{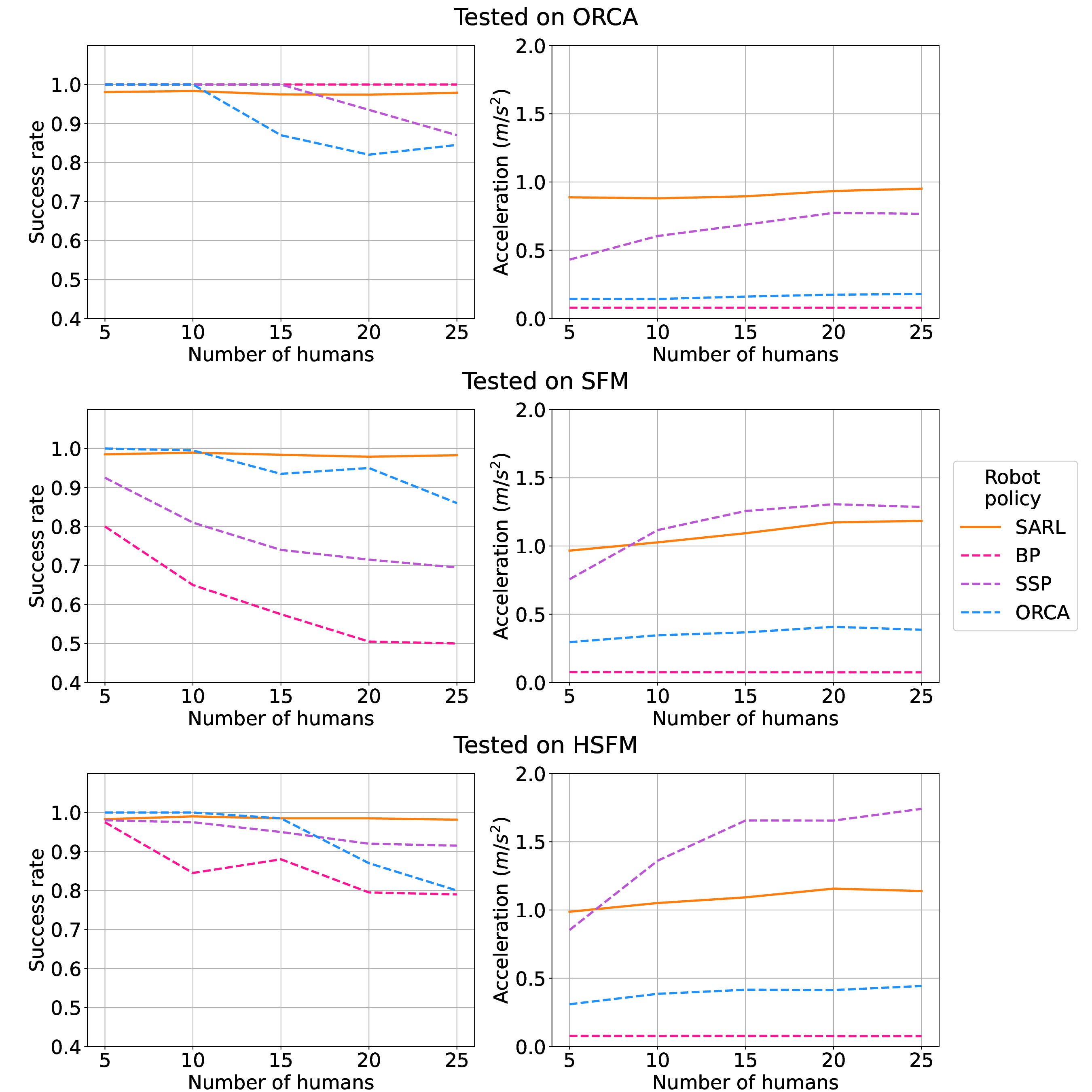}
    \caption{\new{Success rate and average acceleration for SARL and baseline policies (BP, SSP, ORCA) against the number of humans populating the testing environment, for each testing human motion model.} Each baseline curve is averaged over 200 episodes (2 testing scenarios $\times$ 100 trials).}
    \label{fig:baselines_against_sarl_sc_acc}
\end{figure}
\new{The trained policies have been compared to some non trained policies taken from~\cite{mavrogiannis2023core}, used as baselines. Such policies are deliberately simple, with the aim to provide reference values for the considered metrics (e.g., minimum time to goal), and thus to achieve further insights into the behavior of RL-based policies. The first baseline (blind planner, BP) guides the robot at constant speed straight to its target, without taking care of the humans in the scene. The second baseline (simple social planner, SSP) is an algorithm that drives the robot to its target following the shortest path, but stops whenever a human is closer than 20 centimeters. This baseline favors collision avoidance at the expense of navigation efficiency, as opposite to BP. In the third baseline, the robot moves according to the ORCA motion model, which is designed to promote reciprocal avoidance between agents. In Figure~\ref{fig:baselines_against_sarl}, the three mentioned baseline policies are compared to the average results obtained by SARL policies.} It can be noted that the baseline policies (BP, SSP, ORCA) show a worse performance for almost all metrics, as expected. For instance, time to goal is significantly more sensitive to changes in the number of humans for SSP and ORCA, with respect to SARL (while BP achieves the minimum time to goal by construction, regardless of the number of humans in the scene). In addition, the space compliance and minimum distance to humans of all baseline policies are considerably lower with respect to the trained policies, indicating a lack of compliance to social rules. Moreover, even though the robot is visible during tests, baseline policies report a significant lower success rate, especially when the number of humans is increased. Also CADRL and LSTM-RL policies perform significantly better than the considered baselines.

\new{For a further insight, Figure~\ref{fig:baselines_against_sarl_sc_acc} shows the success rate and the acceleration of SARL and baseline policies in the three different testing \HMM. SARL shows an almost 100\% success rate, irrespective of the motion model and number of humans, while all baseline policies exhibit a significant drop in the success rate as the number of humans increases. Moreover, for the baseline policies the SFM seems to be the most challenging testing model, followed by the HSFM and then by ORCA. Concerning the acceleration, it significantly increases for SSP when tested on the two force-based \HMM, providing evidence that SFM and HSFM are more challenging for the robot than ORCA motion model, also in terms of social compliance. By construction, BP keeps the same acceleration in all the testing \HMM, while SARL acceleration is almost unaffected by the motion model and crowd density.}

\new{Overall, the comparison with the considered baselines highlights the ability of RL-based approaches to simultaneously account for both social and navigation requirements.}

\begin{figure}[t]
    \centering
    \includegraphics[width=0.8\columnwidth]{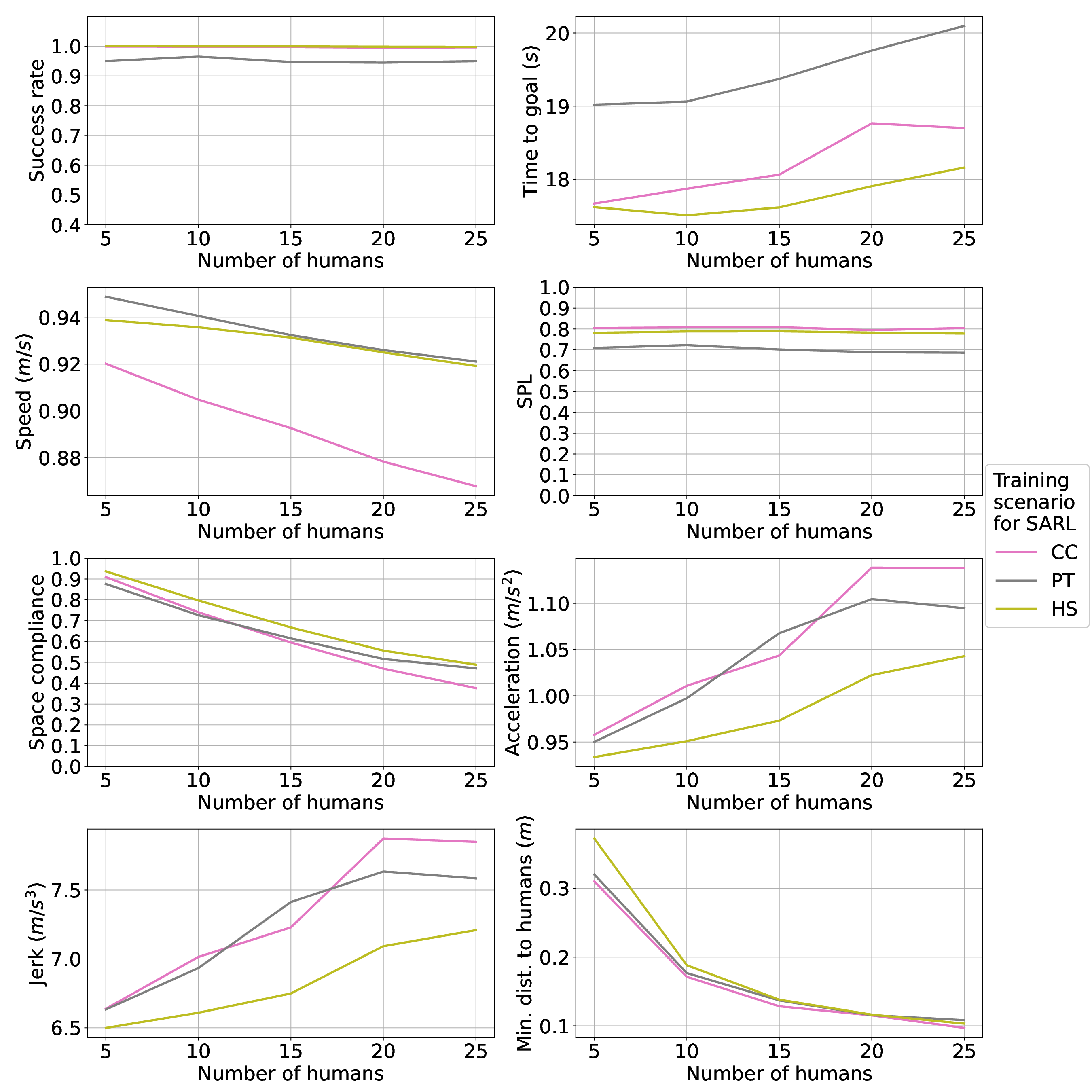}
    \caption{Navigation and social metrics for SARL policies trained in different scenarios against the number of humans populating the testing environment. Each curve is averaged over 1800 episodes (3 training settings $\times$ 6 testing settings $\times$ 100 trials).}
    \label{fig:training_scenarios_metrics}
\end{figure}
\subsection{Comparing scenarios}
In light of the previous results, the effect of different scenarios over training and testing procedures is reported for SARL policies only. Interesting insights can be gained by looking at the performance of the policies trained in different scenarios. 
Figure~\ref{fig:training_scenarios_metrics} depicts the values of the usual metrics for SARL policies, averaged over the 3 different training motion models and the 6 testing settings. The hybrid scenario yields the best performance in terms of time to goal. It also achieves a 100\% success rate, as the CC scenario. Notice that HS and CC have very similar performance in terms of SPL, while PT is slightly worse, mainly due to a lower success rate. Looking at social metrics in Figure~\ref{fig:training_scenarios_metrics}, the hybrid scenario yields the navigation policies with the lowest acceleration and jerk and the largest space compliance. The intuition behind such results is that training on multiple scenarios may be beneficial in terms of both navigation efficiency and social compliance.

\begin{figure}[t]
    \centering
    \includegraphics[width=0.8\columnwidth]{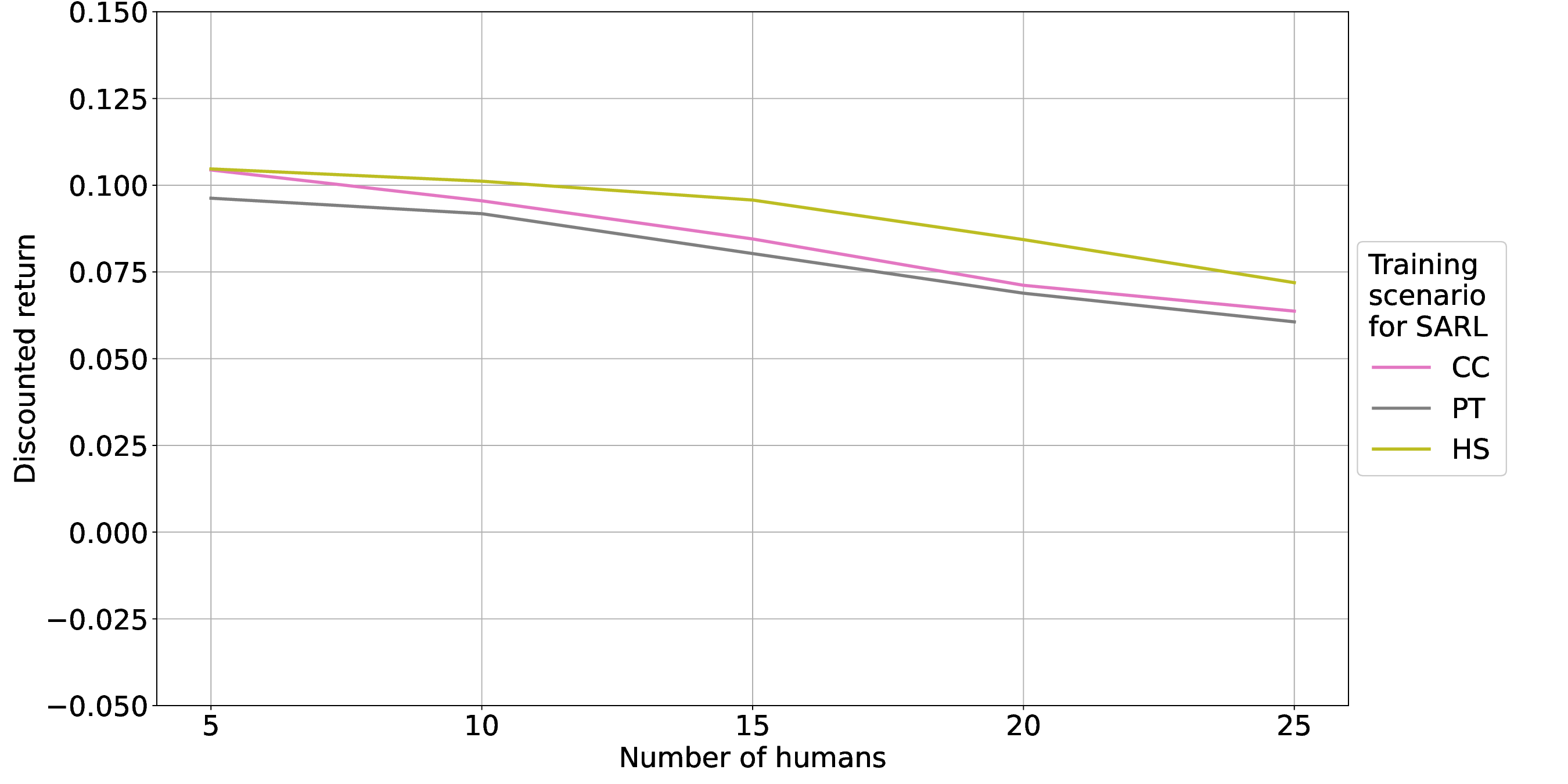}
    \caption{Discounted return (with discount factor $\gamma = 0.9$) for SARL policies trained in different scenarios against the number of humans populating the testing environment. Each curve is averaged over 1800 episodes (as in Figure~\ref{fig:training_scenarios_metrics}).}
    \label{fig:scenarios_return}
\end{figure}

\begin{figure}[t]
    \centering
    \includegraphics[width=0.8\columnwidth]{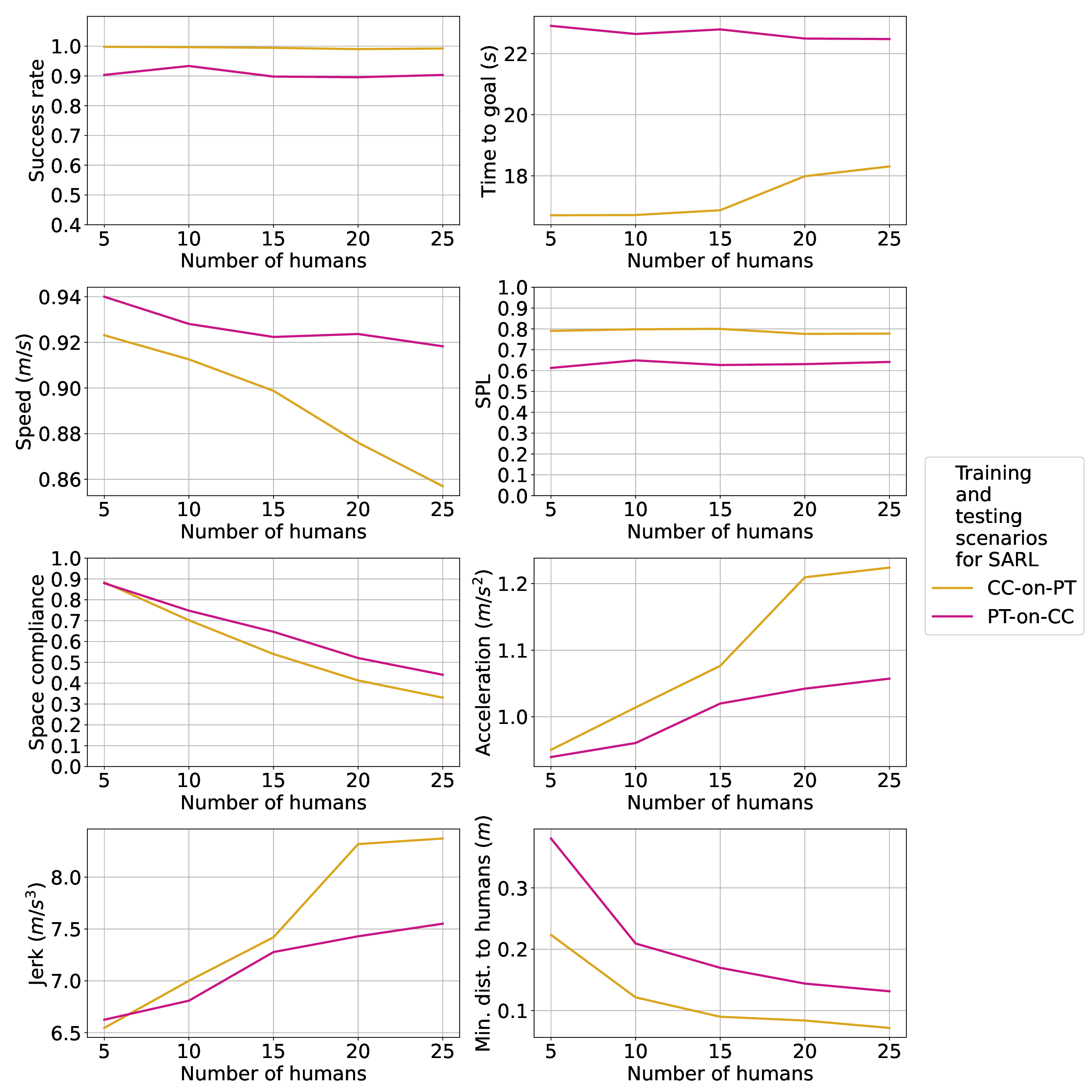}
    \caption{Navigation and social metrics for SARL policies trained in one scenario (either CC or PT) and tested in the other, against the number of humans populating the testing environment. Each curve is averaged over 900 episodes \new{(3 training \HMM\ $\times$ 3 testing \HMM\ $\times$ 100 trials).}}
    \label{fig:cross_scenario}
\end{figure}

Additionally, the discounted return generated by SARL policies trained in different training scenarios has been analyzed. As depicted in Figure~\ref{fig:scenarios_return}, the hybrid scenario is the one yielding trained navigation policies with highest return, indicating again that training on multiple tasks can lead to policies with better performance.
The generalization capabilities of CC and PT as training scenarios have been examined by testing the SARL policies trained in one scenario in a testing setting where the other scenario was employed. Figure~\ref{fig:cross_scenario} reports the results in terms of the usual metrics, for SARL policies trained on the CC scenario and tested on the PT scenario (CC-on-PT), and viceversa (PT-on-CC). Whereas training on CC leads to policies with perfect success rate, low time to goal and high SPL, training on PT yields poorer performance on the CC scenario in terms of navigation metrics. On the other hand, training on PT yields better performance according to all social metrics. Since the parallel traffic scenario is quite human-dense, the robot learns to avoid the crowd by navigating at a certain distance from humans (confirmed by the fact that the SPL is low for policies trained in PT). This can lead to episodes in which, in order to maintain a sufficient distance from humans, the robot fails to reach its goal within the time limit. On the other hand, the circular crossing scenario is less human-dense and, as a result, the robot will be more confident to accomplish its task while crossing the trajectories of humans. In this case, the robot learns to elude collisions with a less safe distance, which is advantageous in terms of navigation metrics.

\begin{figure}[t]
    \centering
    \includegraphics[width=0.8\columnwidth]{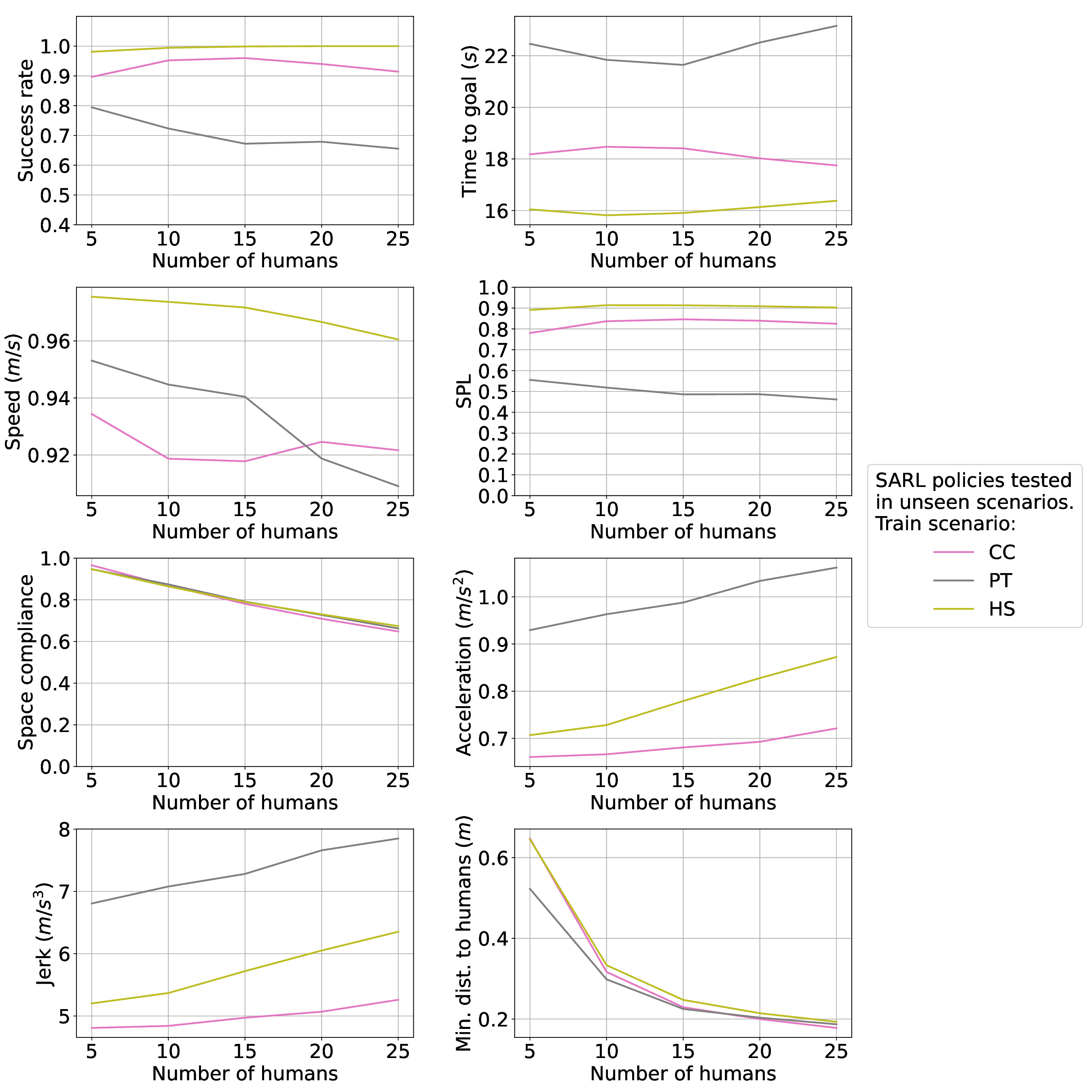}
    \caption{\new{Navigation and social metrics for SARL policies trained in one scenario (CC, PT, HS) and tested in the other different scenarios (PeT, RC, CN), against the number of HSFM-driven humans populating the testing environment. Each curve is averaged over 900 episodes (3 training \HMM\ $\times$ 3 testing scenarios $\times$ 100 trials).}}
    \label{fig:new_scenarios}
\end{figure}

\new{To further assess the generalization capabilities of the trained policies with respect to the testing scenarios, three alternative scenarios, taken from \cite{francis2023principles}, have been used for testing:}
\begin{itemize}
    \item \new{Perpendicular Traffic (PeT): a group of pedestrians continuously flows in one direction (as in PT) and the robot travels in the perpendicular direction aiming to cross the traffic of pedestrians and reach the opposite side of the environment.}
    \item \new{Robot Crowding (RC): humans are randomly placed inside a square of side 7~m and they do not move. The robot is initially placed in the middle of a side and its goal is to reach the mid-point of the opposite side.}
    \item \new{Crowd Navigation (CN): humans are randomly placed inside a circle of radius 7~m and they keep moving within the circle by chasing goals picked at random. The robot is initially on the boundary of the circle and its goal is to reach the point diametrically opposite (this scenario is similar to CC but the pedestrians exhibit a much more irregular and erratic behavior).}
\end{itemize}
\new{All the trained SARL policies have been tested in these scenarios with HSFM-driven humans. Figure \ref{fig:new_scenarios} reports the results achieved by the policies trained in one scenario (either CC, PT or HS) and tested in the PeT, RC and CN scenarios. Notice that none of the testing scenarios has been employed during training. Once again, training in HS yields policies with better performance in terms of navigation efficiency (higher success rate, speed and SPL, lower time to goal) in the previously unseen scenarios. Concerning space compliance and minimum distance to humans, the training scenario does not lead to significant differences, while lower acceleration and jerk can be obtained by training either in CC or HS. Policies trained only in CC or in PT do not demonstrate the same generalization capabilities of those trained in HS. In fact, in the alternative scenarios they achieve a lower success rate with respect to that obtained during tests in CC and PT (see Figure~\ref{fig:training_scenarios_metrics}). Moreover, PT confirms to be a less suitable environment for training.} 

\new{Overall, the study corroborates the idea that the choice of the training scenario does remarkably impact on social navigation and that training in multiple scenarios leads to higher generalization capabilities.}

\subsection{Comparing human motion models for training and testing}
\begin{figure}[t]
    \centering
    \includegraphics[width=\columnwidth]{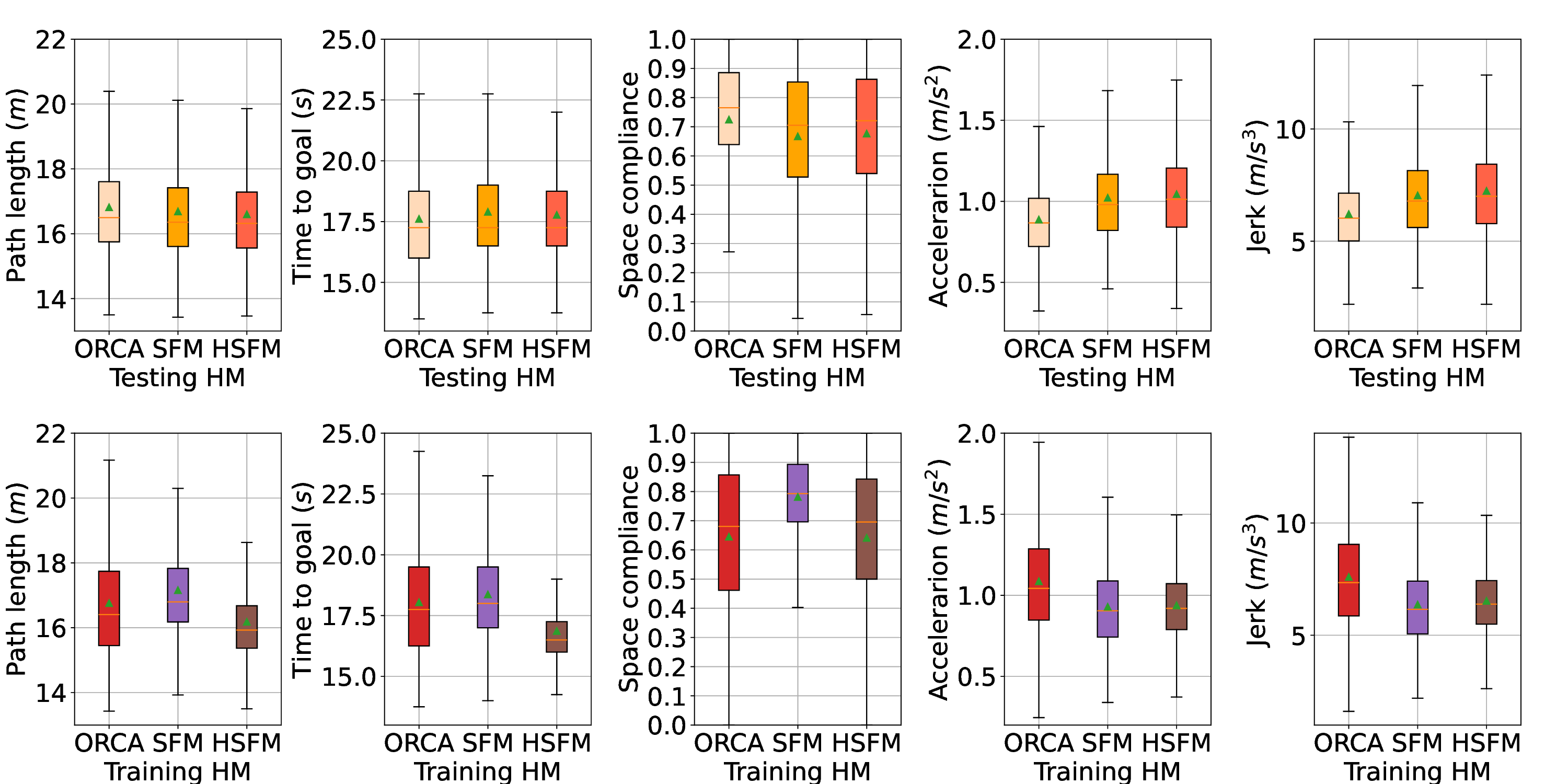}
    \caption{\new{Path length, time to goal, space compliance, acceleration norm and jerk norm for SARL policies trained in the hybrid scenario. The first row shows the metrics for different testing \HMM\ (averaged over the training ones) while the second row reports the metrics cross training \HMM\ (averaged over the testing one). Each box is computed over 3000 testing episodes (3 training/testing \HMM\ $\times$ 2 testing scenarios $\times$ 5 humans densities $\times$ 100 trials). The green triangle of each box represents the mean, the orange line represents the median and the box delimits the interquartile range.}}
    \label{fig:cross_test_env}
\end{figure}

\begin{figure}[t]
    \centering
    \includegraphics[width=0.8\columnwidth]{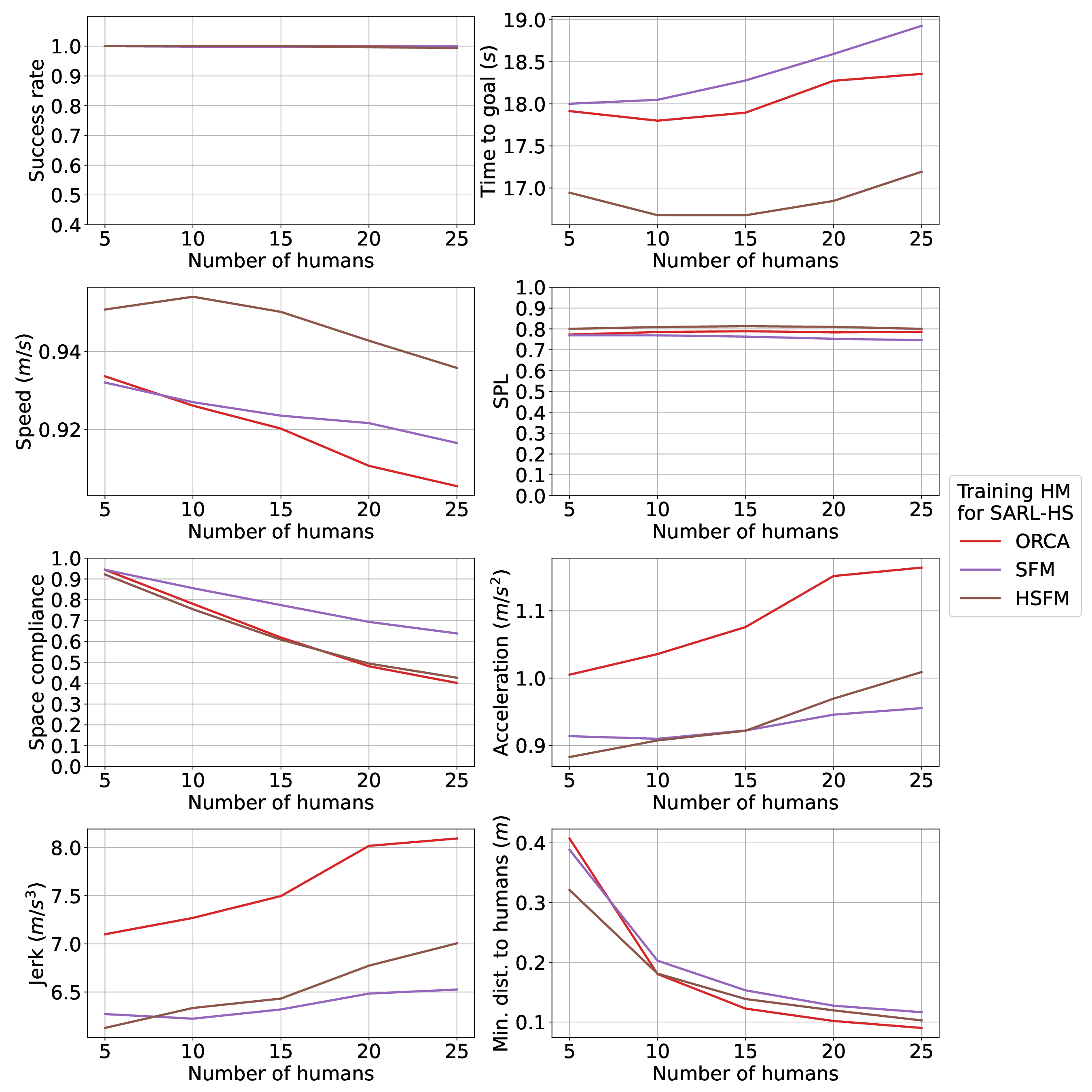}
    \caption{Navigation and social metrics for SARL policies trained in the hybrid scenario and in three different human motion models (ORCA, SFM, HSFM) against the number of humans populating the testing environment. Each curve is averaged over 600 episodes (6 testing settings $\times$ 100 trials).}
    \label{fig:train_env_metrics}
\end{figure}

\begin{figure}[t]
    \centering    \subfloat{\includegraphics[width=0.8\columnwidth]{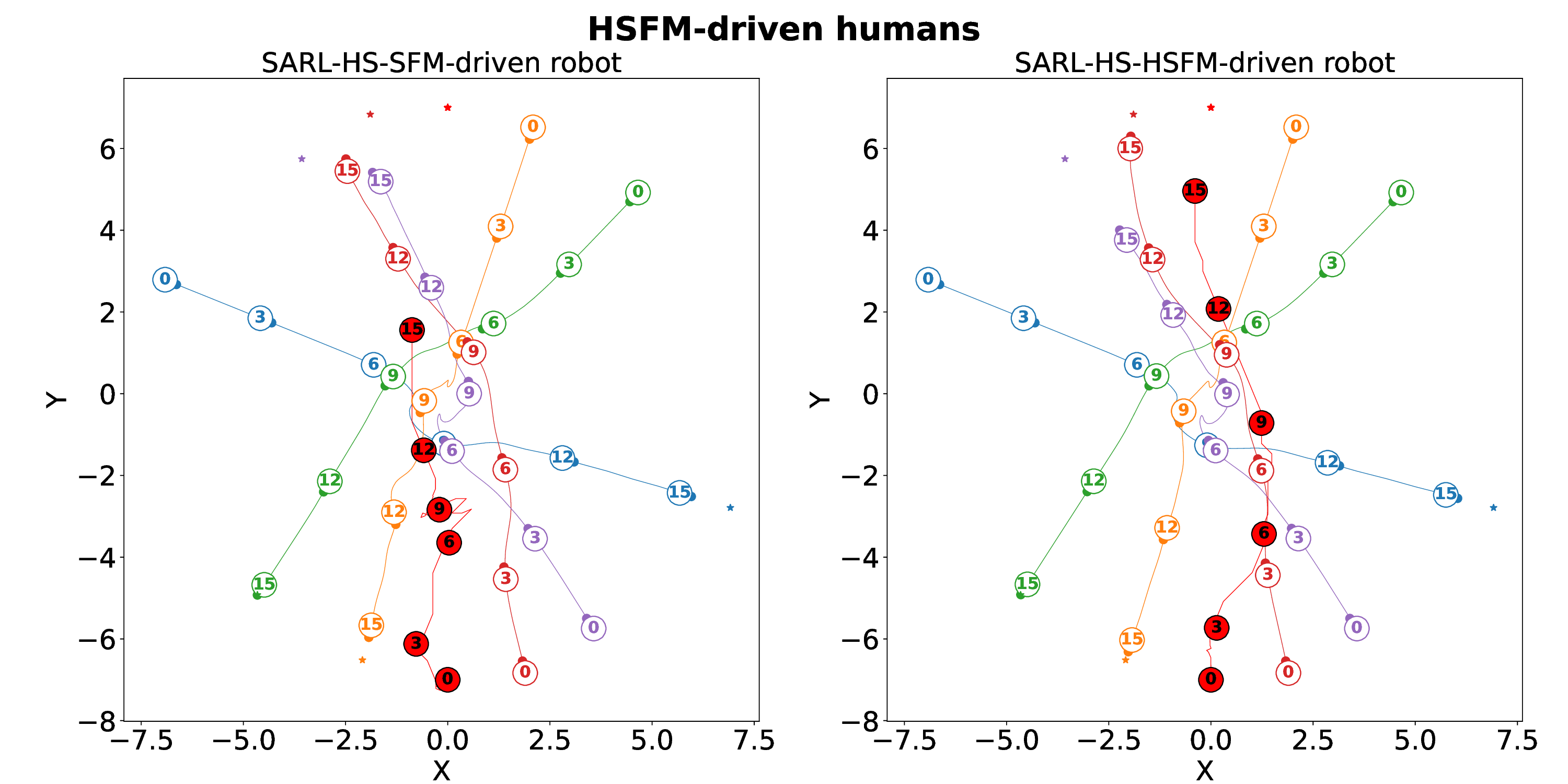}}\\[-2ex] \subfloat{\includegraphics[width=0.8\columnwidth]{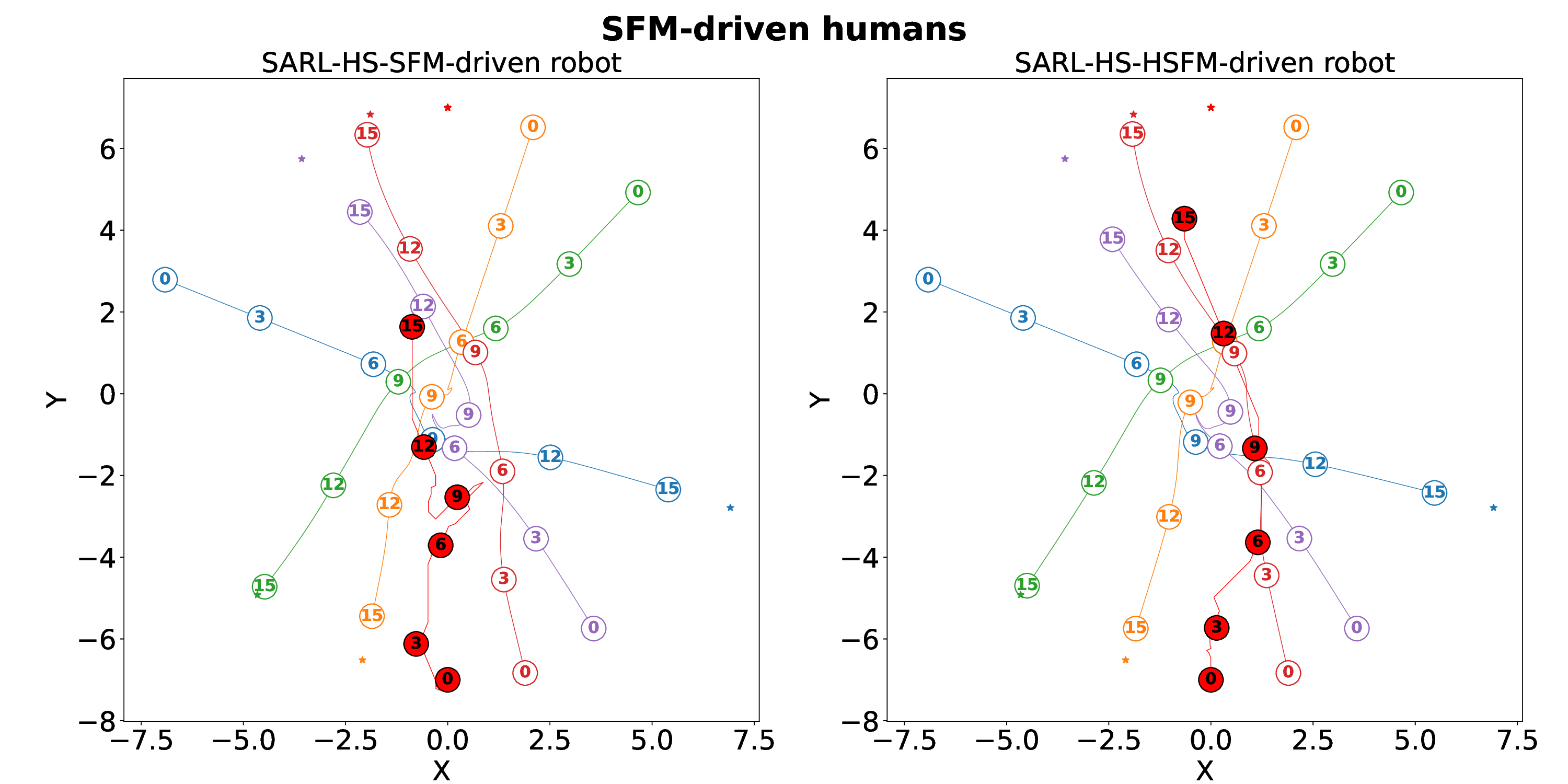}}
    \caption{Trajectories of 5 humans (represented as colored circles) and a robot (red-filled circle) driven by two policies (SARL-HS-SFM and SARL-HS-HSFM) starting from the same initial conditions in the circular crossing scenario. The configurations of all the agents are depicted every three seconds (the time is indicated in the middle of each circle) while the entire path is represented with a segment of the same color of the circle. The star symbols indicate the goals of each agent.}\label{fig:trajectories}
\end{figure}

\begin{figure}[t]
    \centering
    \includegraphics[width=0.8\columnwidth]{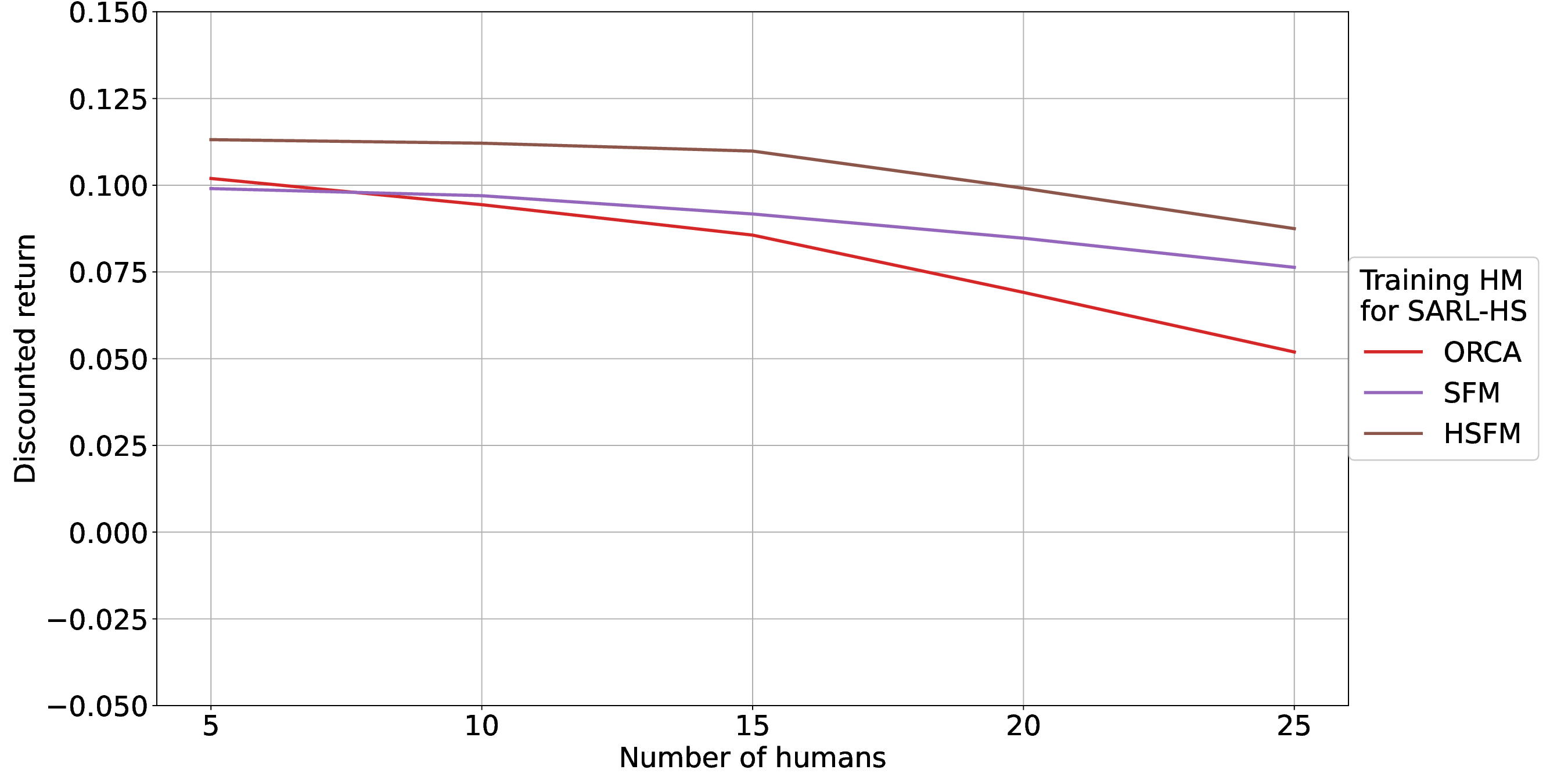}
    \caption{Discounted return (with discount factor $\gamma = 0.9$) for SARL policies trained in the hybrid scenario and in three different human motion models (ORCA, SFM, HSFM) against the number of humans populating the testing environment. Each curve is averaged over 600 episodes (as in Figure~\ref{fig:train_env_metrics}).}
    \label{fig:train_env_return}
\end{figure}

 In order to analyze the impact of the human motion model used in the training and testing settings, the performance of navigation policies trained or tested in ORCA, SFM and HSFM has been evaluated. Considering the previous analysis, we focus on SARL policies trained in the hybrid scenario. Figure~\ref{fig:cross_test_env} reports the boxplots of several navigation and social metrics resulted from testing the SARL policies trained in the hybrid scenario. 
 \new{The first row of plots concerns the statistics for different testing \HMM\ (averaged over the training ones), while the second one covers the statistics for different training \HMM\ (averaged over the testing ones). Note that, in these plots, the SPL is replaced by the path length, because the former is not an episodic metric. It can be observed that the the boxplots of the first row are very similar, indicating that the testing human motion model has a limited impact on the performance of the policy. Instead, much larger differences are observed in the second row, suggesting that the training human motion model affects more significantly the performance. In particular, shorter path lengths and time to goal are observed for policies trained in HSFM, while SFM shows the best results as long as social metrics are concerned.}
\new{The effect of the training \HMM\ (i.e., the human motion model used during the training) with respect to the number of humans in the scene is analyzed in Figure~\ref{fig:train_env_metrics}. Results confirm that training in the HSFM yields navigation policies with better performance in terms of path length and time to goal. As long as social metrics are considered, training in SFM leads to a higher space compliance, with lower acceleration and jerk, and more robustness with respect to the number of humans in the scene. On the other hand, training in ORCA returns policies that are less socially compliant, showing higher acceleration and jerk, especially in crowded environments.}
\new{In order to get further insights into the role of the training human motion model, Figure~\ref{fig:trajectories} reports some episodes in which the robot adopts a SARL policy trained on a hybrid scenario, in either the SFM (SARL-HS-SFM) or the HSFM (SARL-HS-HSFM). The testing setting is the circular crossing scenario, with humans driven either by SFM or HSFM.}
It can be noted that SARL-HS-SFM (plots on the left in Figure~\ref{fig:trajectories}) is more conservative and tends to slow down when crossing human paths in the center of the scene. Instead, SARL-HS-HSFM (plots on the right) is able to cover more distance within the same time frame because it feels comfortable to get closer to humans. This may be due to the fact that training in the HSFM makes the robot more confident, thanks to the smoother motion of HSFM agents. It can also be noted the SARL-HS-HSFM is faster to reach the target and follows smoother paths with respect to SARL-HS-SFM.
This is supported by Figure~\ref{fig:train_env_return}, which shows the discounted return of SARL-HS policies after training in different \HMM, depending on the number of humans. The HSFM is the motion model yielding the policy with highest return.
Given the definition of reward in \eqref{eq:reward}, having the highest return indicates that SARL-HS-HSFM is the policy that reaches the target faster and maintains a minimum separation distance from humans for a longer time, on average.

\begin{figure}[t]
    \centering
    \includegraphics[width=0.8\columnwidth]{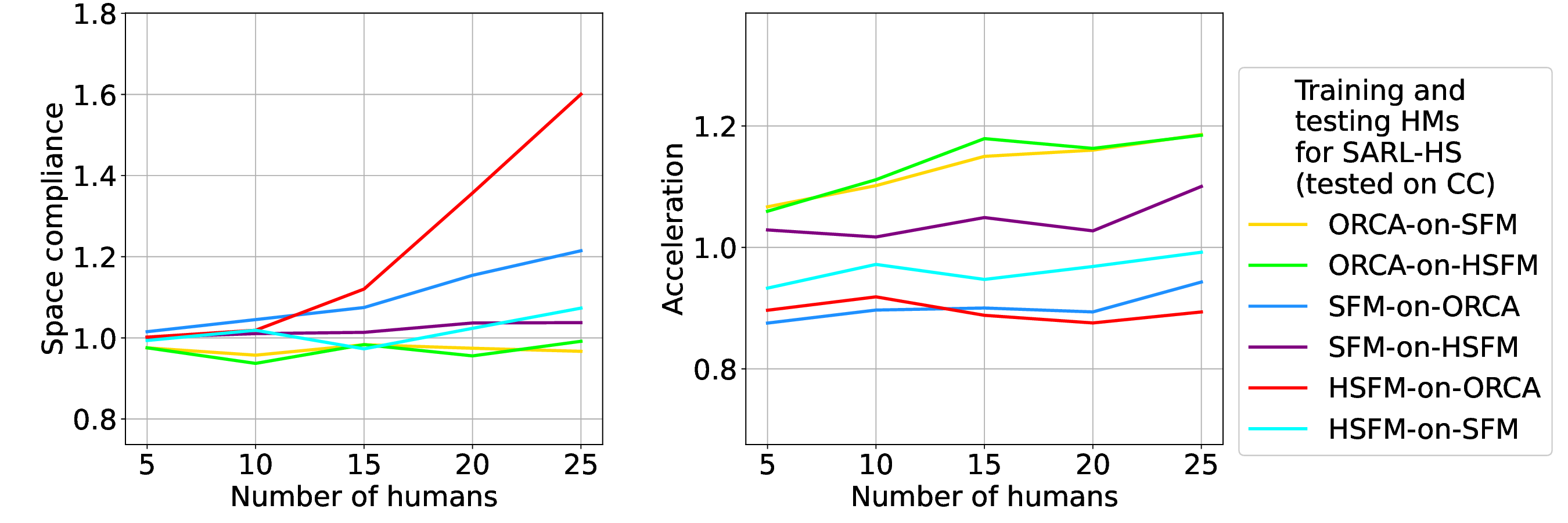}
    \caption{\new{Space compliance and average acceleration for SARL-HS policies trained in one human motion model (ORCA, SFM, HSFM) and tested in a different one, in the circular crossing scenario, against the number of humans populating the testing environment. Each curve is normalized by the performance of the SARL-HS policy trained and tested in the same motion model and is averaged over 100 trials.}}
    \label{fig:cross_env_norm}
\end{figure}
Finally, we evaluate how the human motion model used during training influences the generalization ability of the learned policies. \new{Figure~\ref{fig:cross_env_norm} illustrates the space compliance and average acceleration of SARL-HS policies when applied to a motion model different from the one used for training. These metrics, which are averaged over tests conducted in the CC scenario, are normalized by the performance of the policy trained and tested in the same motion model. For example, the yellow curve labeled ORCA-on-SFM in Figure~\ref{fig:cross_env_norm} represents the performance of SARL-HS-ORCA tested in the SFM, normalized by its performance when tested in ORCA.}

\new{From the space compliance data, it is evident that policies trained in a force-based motion model and tested in ORCA (represented by the SFM-on-ORCA and HSFM-on-ORCA curves in Figure~\ref{fig:cross_env_norm}) demonstrate higher space compliance when tested in ORCA compared to their training motion model (a ratio greater than one). This effect becomes more pronounced as crowd density increases. Regarding average acceleration, the SFM-on-ORCA and HSFM-on-ORCA ratios are less than one, indicating that policies trained in SFM or HSFM generally exhibit lower acceleration when tested in ORCA compared to their training motion model. Conversely, the ORCA-trained policy shows higher acceleration when tested in SFM or HSFM compared to its performance in ORCA (shown by the ORCA-on-SFM and ORCA-on-HSFM curves). Overall, these findings suggest that policies trained in SFM or HSFM demonstrate superior generalization capabilities, particularly in terms of social metrics, compared to those trained in ORCA.}

\subsection{Evaluating scenarios with obstacles}
\new{In this section, the performance of the trained policies is assessed in a case study that includes static obstacles. A variant of the circular crossing scenario, denoted as CCSO (Circular Crossing with Static Obstacles), is introduced to perform the analysis.  Three circular static obstacles  are placed randomly inside the environment, with a total area varying uniformly between 6 and 14~m\textsuperscript{2}. The initial human and robot positions and their goals are set as in the standard CC scenario.}

\begin{figure}[t]
    \centering
    \includegraphics[width=0.8\columnwidth]{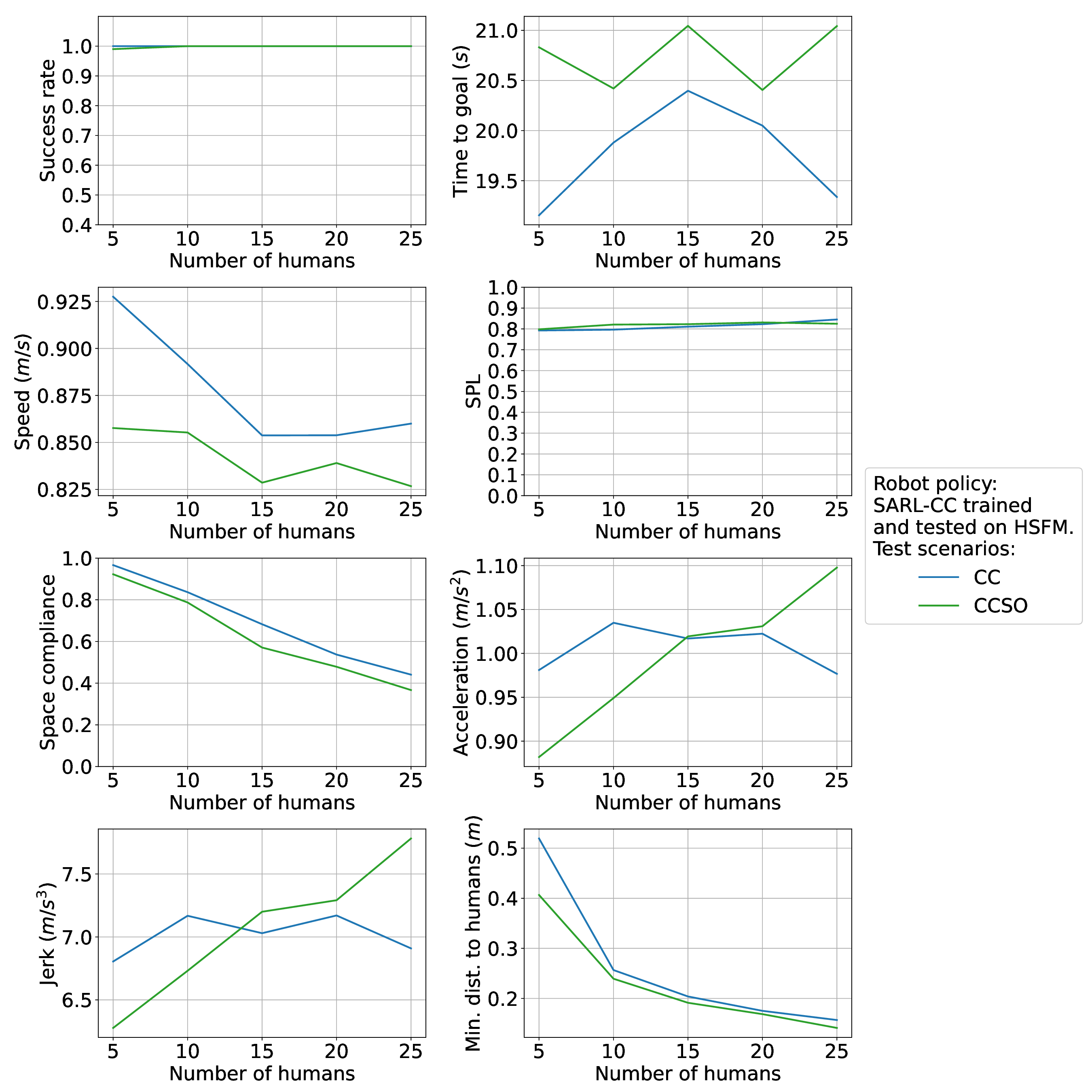}
    \caption{\new{Navigation and social metrics for the SARL policy trained in the circular crossing scenario without static obstacles and tested in the circular crossing with (CCSO) and without (CC) static obstacles. In both training and testing humans are HSFM-driven. Each curve is averaged over 100 trials.}}
    \label{fig:static_obstacles}
\end{figure}

\new{Figure \ref{fig:static_obstacles} reports the performance of the SARL policy trained in the CC scenario without obstacles and tested both in CC and in CCSO. In both the training and the testing phase humans are HSFM-driven. When testing in CCSO, the navigation policy shows a 100\% success rate although it has been trained in an environment without static obstacles. From a navigation viewpoint, the time to goal and speed slightly get worse due to the presence of obstacles. On the other hand, in terms of social metrics the performance does not deteriorate significantly.}

\new{Overall, this case study indicates that the trained policies are robust with respect to the presence of static obstacles in the workspace.}

\subsection{Testing robot policies with noisy measurements}
\new{In the analysis conducted so far, tests have been carried out in noiseless environments, i.e., it is assumed that the true joint state $\mathbf{s}^{jn}$ is available to the robot (notice that this is a common assumption in works as \cite{chen2017decentralized,everett2018motion,SARL}). In this section, the effect of noisy conditions is evaluated.}

\begin{figure}[t]
    \centering
    \includegraphics[width=0.8\columnwidth]{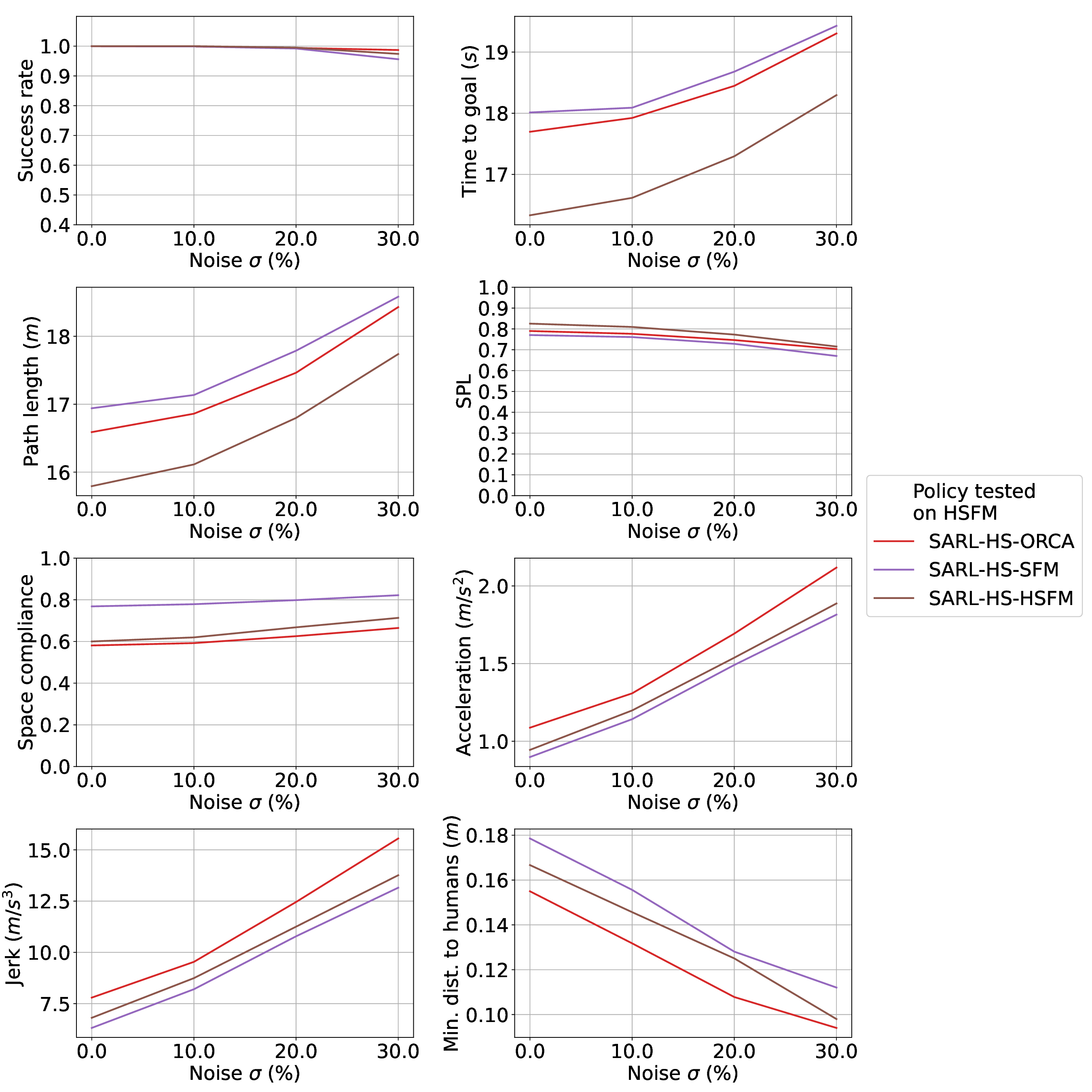}
    \caption{\new{Navigation and social metrics for SARL policies, trained in the hybrid scenario in three different HMs (ORCA, SFM, HSFM) and tested with noisy measurements and 15 HSFM-driven humans, against the noise standard deviation. Each curve is averaged over 200 episodes (2 testing scenarios $\times$ 100 trials).}}
    \label{fig:noisy_tests}
\end{figure}

\new{In order to simulate measurement errors, the observed human position and velocity relative to the robot are corrupted by a zero-mean Gaussian noise, whose standard deviation $\sigma$ is a given percentage of the nominal measurement. Figure~\ref{fig:noisy_tests} reports the results obtained by SARL policies trained in the hybrid scenario and tested with noisy data in the CC and PT scenarios with 15 humans moving according to the HSFM. The results are reported for the different training \HMM\ against different levels of $\sigma$. As shown, despite being trained in a noiseless environment, the policies achieve a very high success rate also with significant noise. The space compliance is essentially insensitive to noise, while the other metrics drop as noise increases. Nevertheless, the overall performance remains more than acceptable even for extremely high noise levels (up to 30\%).}

\new{On the whole, the results show that SARL policies are remarkably robust to the presence of noise in the human motion measurements.}

\subsection{Discussion}
\new{The simulation campaign presented in this study provides valuable insights into the role of human motion models for RL-based navigation policies in crowded environments. The evaluated policies demonstrate solid performance, being able to drive the robot to the target most of the time and to cope with the presence of humans in the operating space. 
Among the RL policies tested, SARL emerges as the most effective, offering a good balance between navigation and social metrics. It achieves near-perfect success rates, the shortest time-to-goal, and the most efficient path lengths, while maintaining low acceleration and jerk. Additionally, SARL displays remarkable robustness to varying crowd densities, which corroborates its practical viability for real-world applications.}

\new{Training policies across diverse scenarios has proven advantageous, leading to more robust navigation strategies adaptable to different environments. Training in multiple scenarios leads to better generalization capabilities even with respect to previously unseen testing scenarios. Moreover, the performance of SARL policies remains consistent when tested in scenarios with static obstacles, even if they were not present in the training setting. This emphasizes the importance of considering a variety of scenarios to develop versatile and resilient navigation systems.}

\new{The human motion model adopted during training plays a major role. 
Of the three considered human models, ORCA leads to the poorest performance in terms of social metrics. Navigation policies trained in ORCA result in trajectories with the worst space compliance, acceleration and jerk. This is especially true in the presence of high crowd density. In this respect, SFM and HSFM as training motion models provide better results. As a general comment, training in HSFM favors navigation aspects (shorter path length and time-to-goal), whereas SFM is better for social aspects (larger space compliance, lower acceleration and jerk). Training in SFM provides policies which are more robust with respect to the number of humans populating the environment.
Moreover, looking at the resulting trajectories, it turns out that HSFM favors faster and smoother paths than SFM, while ensuring a reasonable minimum distance to humans. This nice property was observed regardless of the human motion model adopted during the tests. It has also been verified that RL-based policies exhibit only a minor performance degradation when the measurements of the pedestrian positions and velocities are corrupted by noise.}
 
\new{With respect to the simulation study presented in \cite{mavrogiannis2023core}, the analysis carried out in this work confirms several outcomes. In particular, the fact that testing on the ORCA motion model may not be fully informative about the performance of the navigation algorithms. This is probably due to the cooperative nature of ORCA which might not be representative of real crowd behavior. Force-based models introduce a higher level of complexity and provide a more challenging and realistic environment, enhancing the practical applicability of the learned policies. The extensive simulation campaign involving different scenarios and pedestrian densities has further demonstrated the importance of the human motion model in the training phase, rather than in the testing one. It has also shed light on the superiority of navigation techniques based on reinforcement learning with respect to non-trained baselines, such as SSP or ORCA, especially in terms of robustness with respect to the crowd density. Analyzing the performance of the baselines reveals that the most difficult motion models, in terms of success rate, are SFM, HSFM, and ORCA, in decreasing order. This suggests that training in SFM is particularly tough, leading to more cautious policies that emphasize social compliance. Training in HSFM is also challenging but less so than SFM, resulting in more efficient navigation paths. Conversely, ORCA appears to be less complex as a training motion model, so the robot struggles with more complex behaviors, leading to poorer performance in both social and navigation metrics. Summing up, HSFM can be seen as a good compromise between the more navigation-oriented design of ORCA and the socially-oriented features of SFM.
On the whole, the results of the present study seem to suggest that, although not fully realistic, the existing human motion models are rich enough to provide challenging training and testing environments for learning-based navigation techniques, provided that they are used in combined fashion. A possible good practice may involve training on multiple scenarios in HSFM and then several testing sessions on ORCA and SFM, including novel scenarios not used in the training phase. Additionally, performance evaluation should take into account both navigation and social metrics, as well as different crowd density levels.}


\section{Conclusions}\label{conclusions}
An extensive comparative study of human motion models for learning-based robot navigation policies has been carried out. Three RL algorithms (CADRL, LSTM-RL and SARL) have been trained and tested using three different approaches to human modeling (ORCA, SFM and HSFM). The simulation campaign reveals that RL-based navigation policies generally perform well, handling human presence effectively and outperforming naive methods in both efficiency and social compliance. Among the tested policies, SARL offers the best overall performance, with high success rates, efficient paths, and low acceleration, while being less affected by crowd density. 
Concerning human motion models, the choice is more critical during training than testing. ORCA leads to poorer social metrics especially in dense crowds,  whereas SFM and HSFM offer better outcomes, with SFM favoring social metrics and HSFM favoring navigation efficiency. SFM-trained policies are more robust to crowd size, while HSFM-trained policies ensure faster, smoother paths. 

\new{Although velocity-based and force-based human motion models provide a robust foundation for training and testing navigation policies, it is clear that they also have serious limitations. Human movements are far more complex than those of circle-shaped agents. Moreover, the interaction between a pedestrian and a robot may be much more intricate than reciprocal avoidance. The RL-based navigation policies considered in this work assume that the state of the pedestrians is available to the robot, possibly corrupted by noise, but in the real world it might be necessary to estimate such quantities online. These considerations suggests that the results of the present study are just a first step towards a full understanding of the role of human motion models in social robot navigation, and much work remains to do to  fully address the complexities of real-world environments. 
Among the many possible lines of development, one notable example is the construction and validation of models based on data coming from real crowd behaviors. One may use such data to suitably tune the parameters of velocity-based or force-based motion
models. A thorough comparison with machine learning approaches for data-driven prediction of pedestrian trajectories would further validate the ability of such models to reproduce complex crowd behaviors. More challenging scenarios for training and testing the navigation algorithms should also be considered, including complex navigation tasks or featuring various types of human-robot interactions. Enhanced adherence to real-world contexts may be achieved by employing the considered human motion models as low-level dynamic engines within more realistic software platforms for the simulation of pedestrians motion. To further broaden the study, navigation pipelines not based on machine learning (such as those relying on rapidly exploring random trees or model predictive control) may be addressed. The presence of multiple robots in the scene is another setting that deserves deeper investigations,  which can benefit from an extensive use of human motion models.
}

\bibliographystyle{ieeetr}
\bibliography{refs}

\end{document}